\documentclass[fleqn,usenatbib]{mnras}

\usepackage{newtxtext,newtxmath}

\usepackage[T1]{fontenc}
\usepackage{ae,aecompl}
\usepackage{graphicx}	
\usepackage{amsmath}	

\title[Planet formation in intermediate-separation binaries]{Planet formation in intermediate-separation binary systems}

\author[O.Pani\'c et al., Planet formation in binary systems]{
O. Pani\'c$^{1}$\thanks{E-mail: o.panic@leeds.ac.uk},
T. J. Haworth$^{2}$, 
M. G. Petr-Gotzens$^{3,4}$,
J. Miley$^{1,14}$,
M. van den Ancker$^{3}$, 
\newauthor
M. Vioque $^{1,14}$,
L. Siess$^{5}$,
R. Parker$^{6}$,
C. J. Clarke$^{7}$, 
I. Kamp$^{8}$, 
G. Kennedy$^{9}$, 
\newauthor
R. D. Oudmaijer$^{1}$, 
I. Pascucci$^{10}$, 
A. M. S. Richards$^{11}$, 
T. Ratzka$^{12}$, 
C. Qi$^{13}$ 
\\
$^{1}$School of Physics and Astronomy, University of Leeds, Leeds LS2 9JT, UK\\
$^{2}$Astronomy Unit, School of Physics and Astronomy, Queen Mary University of London, London E1 4NS, UK \\
$^{3}$ European Southern Observatory Headquarters, Karl-Schwarzschild-Strasse 2, D-85748 Garching bei M\"unchen \\
$^{4}$ Universit\"ats-Sternwarte, Ludwig-Maximilians-Universit\"at M\"unchen, Scheinerstr 1, D-81679 M\"unchen, Germany \\
$^{5}$ Institut d'Astronomie et d'Astrophysique, Universit\'e Libre de Bruxelles (ULB), CP 226 - 1060, Bruxelles, Belgique \\
$^{6}$ Department of Physics and Astronomy, The University of Sheffield, Hicks Building, Hounsfield Road, Sheffield, S3 7RH, UK \\
$^{7}$ Institute of Astronomy, University of Cambridge, Madingley Road, Cambridge, CB3 0HA, UK \\
$^{8}$ Kapteyn Astronomical Institute, University of Groningen, PO Box 72, NL-9700 AB Groningen, the Netherlands \\
$^{9}$ Department of Physics, University of Warwick, Gibbet Hill Road, Coventry CV4 7AL, UK \\
$^{10}$ Lunar and Planetary Laboratory, The University of Arizona, Tucson, AZ 85721, USA \\
$^{11}$ Jodrell Bank Centre for Astrophysics, School of Physics and Astronomy, University of Manchester, Manchester M13 9PL, UK \\
$^{12}$ Institute for Physics/IGAM, NAWI Graz, Karl-Franzens-Universit\"at, Universit\"atsplatz 5/II, A-8010 Graz, Austria \\
$^{13}$ Harvard-Smithsonian Center for Astrophysics, 60 Garden Street, Cambridge, MA 02138, USA \\
$^{14}$ Joint ALMA Observatory, Alonso de C\'ordova 3107, Vitacura 763-0355, Santiago, Chile \\
}

\date{Accepted XXX. Received YYY; in original form ZZZ}

\pubyear{2020}

\begin{document}
\label{firstpage}
\pagerange{\pageref{firstpage}--\pageref{lastpage}}
\maketitle
%
%
\begin{abstract}
We report the first characterisation of the individual discs in the intermediate separation binary systems KK~Oph and HD~144668 at millimetre wavelengths. In both systems the circum-primary and the circum-secondary discs are detected in the millimetre continuum emission, but not in $^{13}$CO nor C$^{18}$O lines. Even though the disc structure is only marginally resolved, we find indications of large-scale asymmetries in the outer regions of the primary discs, most likely due to perturbation by the companion. The derived dust masses are firmly above debris disc level for all stars. The primaries have about three times more dust in their discs than the secondaries. In the case of HD~144668 the opacity spectral index of the primary and secondary differ by the large margin of 0.69 which may be a consequence of the secondary disc being more compact. Upper limits on the gas masses imply less than 0.1~M$_{\textrm{jup}}$ in any of these discs, meaning that giant planets can no longer form in them. Considering that there have been no massive gas discs identified to date in intermediate separation binaries (i.e., binaries at a few hundred au separation), this opens space for speculation whether their binarity causes the removal of gas, with tidal interaction truncating the discs and hence shortening the accretion timescale. More systematic studies in this respect are sorely needed.
\end{abstract}




\section{Introduction}
The need to explain the diverse properties of exoplanets being discovered \citep{2015ARA&A..53..409W} is continuing to drive research into protoplanetary disc evolution and planet formation \citep[there are a number of reviews of observational and theoretical work, e.g.][]{2011ARA&A..49...67W, 2015PASP..127..961A, 2016PASA...33...53H, 2017RSOS....470114E}. The main focus in theoretical research into this problem considers the formation of planets from isolated discs. However there are environmental factors that can play a role, one of which is stellar multiplicity. 

A large fraction of Solar-mass stars form in binaries \citep{1991A&A...248..485D}.
This fraction is even higher for stars with masses above 2~M$_{\odot}$ \citep[70\%,][and references therein]{2013ARA&A..51..269D}. 
In particular, the young, intermediate-mass stars with circumstellar material (known as Herbig Ae/Be stars)
are found to form predominantly as multiples as attested by a number of studies \citep[e.g.][]{2006MNRAS.367..737B, 2015Ap&SS.355..291D}.
The separation of the binary pair is expected to have a significant impact on the disc evolution. For binaries on close orbits (short periods) their mutual ``circumbinary disc'' is subject to enhanced torques from the binary potential that result in an asymmetric, elliptical, often precessing disc \citep[e.g.][]{10.1046/j.1365-8711.2001.04011.x, 10.1093/mnras/stu663, 2017MNRAS.469.2834O}. Circumbinary discs also have dynamically excavated inner holes \citep[e.g. GG Tau][]{2014Natur.514..600D}
and can have warped or completely misaligned components \citep[e.g.][]{1977MNRAS.181..441P, 1983MNRAS.202.1181P, Kennedy2019}. For binaries with larger separations tidal truncation of the individual discs can physically constrain the range of orbital distances within which planets can form. This occurs when torques from the binary extract angular momentum more effectively than the extraction of angular momentum by viscous evolution in the disc itself \citep{1980ApJ...241..425G}. This more effective removal of angular momentum can also promote accretion. Coronographic imaging has revealed that G-K-M type stars have a lack of substellar companions in the range 75-300~au \citep{2004AJ....127.2871M}. 
An external binary partner can also induce asymmetries such as spirals in the primary disc \citep{2018ApJ...869L..44K}.

Exoplanet statistics suggest that binarity could promote the formation of certain types of planets. In fact, \citet{Desidera2006PropertiesSeparation} established that short-period massive planets are preferentially found around stars which do have a companion at separation of less than a few hundred au. In a more recent study, \cite{2019MNRAS.485.4967F} found a high incidence of such companions for stars hosting giant planets with masses $\ge$3-4~M$_{jup}$ and orbits within 1~au. This supports the theoretical expectation that such planets are formed via gravitational instability, with the companion triggering the planet formation and facilitating the inward migration of these planets \citep[e.g.][]{2008A&A...480..879S, 2010ApJ...708.1585K, 2011MNRAS.417.1928F, 2017MNRAS.470.2517H}

\citet{Zucker2002ONPLANETS} show that massive planets have a shorter, one-day period when orbiting stars which have a binary companion, and a longer, three-day period around single stars.
They show that planet-host binary stars, including binary separations up to 1000~au, have an anticorrelation of planet mass with the period of the planet, while a strong correlation is present for single stars. \cite{2016AJ....152....8K} also found that close stellar binaries can suppress planet occurrence. There is hence growing evidence that  binarity influences the frequency and types of planets formed, which is important for the predominantly multiple Herbig Ae stars.

A crucial insight into planet formation from discs around intermediate-separation binaries (a few hundred au) can be provided by observations in the millimetre wavelength regime, which can yield information on the disc masses and  reveal large scale structures. In one such study of Taurus multiple systems, \citep{2012ApJ...751..115H} showed that there is a marked transition around 100-300~au projected separation where the more widely separated pairs had significantly larger disc masses compared to the pairs with projected separations of less than 100~au. This shows that stellar binarity, and orbital separation in particular, are strongly related to disc mass and hence the ability of discs to form planets. In fact, close-separation binaries are found to have lower disc masses \citep{1994ApJ...429L..29J, 2005ApJ...631.1134A, 2012ApJ...751..115H, 2018ApJ...863...61L} and sizes \citep{2019A&A...628A..95M}. Furthermore, a survey of excess emission in a large sample of main sequence stars with binary separations of less than 100~au \citep{2012ApJ...745..147R} has shown lower debris disc masses in such systems, with possible implications for rocky planet formation and evolution.

\cite{2017ApJ...851...83C} surveyed discs towards Ophiucus with ALMA 870~$\mu$m continuum observations. They found that the binaries had lower disc masses/radii. They also found that their dust discs were less extended than the theoretical truncation radius, but considering that these observations probe preferentially millimetre dust prone to drift, observational tracers of the extent of the gaseous component in discs are needed to test whether these discs are truncated by the companion stars.

In this paper we present ALMA images\footnote{This paper makes use of the following ALMA data: 2013.0.01600.S and 2013.1.00220.S. ALMA is a partnership of ESO (representing its member states), NSF (USA) and NINS (Japan), together with NRC (Canada), MOST and ASIAA (Taiwan), and KASI (Republic of Korea), in cooperation with the Republic of Chile. The Joint ALMA Observatory is operated by ESO, AUI/NRAO and NAOJ.} 
of two intermediate-separation binary systems, KK~Oph and HD~144668. Each star in these systems hosts its own protoplanetary disc, and there is no evidence for larger, common, circumbinary material. 

The relatively low stellar densities in Lupus III and Ophiuchus mean that the systems at separations of a few hundred au are in the hard regime, meaning that they can be considered as having evolved in a dynamical isolation from the other cloud members. This, together with their mass ratio of 1-2 strongly supports the idea that they have formed as binaries rather than having been formed through capture \citep{2010MNRAS.404.1835K}. The fact that each star, the primaries and the secondaries, are surrounded by their own circumstellar discs, as shown by our mm observations, provides further support that each pair studied here consists of coeval stars.

\citet{2014MNRAS.437.1216D} find that over 40\% of A type stars are binaries with separations $>$50~au (limit of imaging surveys), with a peak of the distribution at 200-300~au separations. This means that the systems we investigate here are very frequent outcomes of A-type star formation, which is important given the result that giant planets are most frequent around stars $\sim$2~Msun (A-type) \citep{2015A&A...574A.116R}. Another factor of interest for understanding disc evolution and planet formation in A-star binaries is that the two systems discussed here have markedly different ages, with HD~144668 at 2.8$\pm$1~Myr and KK~Oph at 8.0$\pm$2~Myr \citep{Meeus2012AstrophysicsDiscs}.

In this paper, we carry out characterisation of the discs and discuss how their properties may have been influenced by the binary nature of their stellar hosts. Overall we aim to understand the evolution of such discs and what impact binarity may have on their planet forming potential, which might help to explain the observed properties of exoplanets in binary systems.

\section{Pre-main sequence stars KK~Oph and HD~144668, and their companions}

\subsection{KK~Oph}

KK~Oph (Hen 3-1346, AS 220) is an A-type pre-main sequence star associated with an infrared excess characteristic of a protoplanetary disc \citep{1960ApJS....4..337H,Hillenbrand1992HerbigDisks,2005AJ....130..815H}, and stellar photometry which is highly variable. 
KK~Oph forms a binary system with KK~Oph~B at projected separation of 1$\farcs$6 \citep{ 1997A&A...318..472L, 1997ApJ...481..392P}.  Spatially resolved spectroscopy and optical imaging by \citet{Carmona2007OpticalStars}
has shown that the secondary is a mid G-type T Tauri star. They also provide a very detailed study of the stellar properties, assuming a distance of 140~pc, which has more recently been found to be greater (see section Sect.~\ref{opt}). In this paper, we adopt the updated distance of 221.1$^{+12.4}_{-10.4}$~pc, but note that there are great uncertainties on the distance estimates.

KK~Oph is seen in the proximity of the edge of the dark cloud Barnard 59 and of the Sco OB2-2 complex.  However, at a distance of 221~pc, KK~Oph is unlikely to belong to either, as also indicated by \citet{2005AJ....130..815H}.

Optical long-baseline interferometric observations have confirmed the existence of disc material at sub-au scales, in both the primary and the secondary,
in a disk-like geometry seen at a relatively high inclination \citep[$i$ $\approx$ 70$^\circ$][] 
{2004A&A...423..537L, 2013A&A...551A..21K}. \cite{2013A&A...551A..21K} find the companion KK~Oph~B to lie nearly along the direction of the major axis of the disc around KK~Oph~A. Both binary components 
appear to still be actively accreting material, at rates of 
5 $\times$ 10$^{-7}$ M$_\odot$~yr$^{-1}$, and 
1.2 $\times$ 10$^{-8}$ M$_\odot$~yr$^{-1}$, respectively \citep{2006A&A...456.1045B, 2006A&A...459..837G}. \citet{Fairlamb2015ARates} derives a lower rate of $<$10$^{-8}$ M$_\odot$~yr$^{-1}$, possibly due to variable accretion. Low accretion rates are indicative of a low gas mass of the disc \citep{1998ApJ...495..385H, 2016A&A...591L...3M}. In Submillimeter Array (SMA) observations by \citet{Meeus2012GASPSDiscs} at 1.2~mm wavelength, the total continuum emission from the binary system is detected at 36~mJy. The main $^{12}$CO isotopologue emission was detected in Atacama Submillimeter Telescope Experiment (ASTE) observations, \citet{Hales2014AFORMATION} report 0.6~Kkms$^{-1}$ line intensity of the $^{12}$CO J$=$3--2 line.

\subsection{HD~144668}

HD~144668 is an A type star also known as V856~Sco \citep{Hillenbrand1992HerbigDisks,Alecian2012AMeasurements}. The star belongs to the Lupus star forming region and has an SED characteristic of a settled protoplanetary disc \citep{Meeus2012GASPSDiscs}. In fact its fractional near- to far-IR excesses are about 10 times lower than for KK~Oph \citep{Pascual2016ThePhotometry}. Gaia distance to HD~144668 is 161$\pm$32~pc as discussed in Sect.~\ref{opt}.

Almost nothing, except for the approximate J2000 coordinates 16h08m34.4s -39d06m20s is reported about the nearby secondary star **~DUN~199B (HD~144668~B hereafter) at the projected separation of 1.5$\arcsec$ (SIMBAD) in the existing literature.

\cite{Ansdell2016ALMAMasses} report a flux of 55.81$\pm$0.36~mJy at 0.89~mm for the HD~144668 system, and upper limits on $^{13}$CO and C$^{18}$O lines from which they derive an upper limit on the gas mass in the system of 1~M$_{\textrm{jup}}$, but adopt a somewhat larger distance of 200~pc compared to the Gaia DR2 distance of 161$\pm$32~pc. A follow-up analysis by \cite{Miotello2016LupusDepletion} lowered this limit to 0.3~M$_{\textrm{jup}}$. Neither papers discussed the star **~DUN~199B at the projected separation of 1.5$\arcsec$ (SIMBAD), and its associated millimetre emission. We revisit these data, extracting the fluxes separately for the primary and the secondary from the ALMA image. 

\section{Observations}

\subsection{ALMA observations}
The KK Oph and HD144668 systems were observed as part of the ALMA  programme 2015.1.01600.S (P.I. Pani\'c) in Band 6 (1.3~mm) at an angular resolution of 0$\farcs$5 (Tab.~\ref{tab1}) on 2016 May 15 and June 11, respectively. Baseline ranges for both observations were between 14 and 640~m, using 41 of ALMA's 12~m antennas for KK Oph and 38 antennas for HD144668. The spectral setup included $^{13}$CO and C$^{18}$O J$=$2--1 lines with rest frequencies of 220.3986841281 and 219.5603541 respectively (LAMDA database, https://home.strw.leidenuniv.nl/~moldata/CO.html), at velocity resolutions of 0.167 and 0.333~km/s as well as two continuum spectral windows of aggregate bandwidth of 3.75~GHz. 

For KK Oph, J1700-2610 was used as phase calibrator and J1733-1304 as flux calibrator. For HD~144668 the flux calibrator was Titan and the phase calibrator J1610-3958. Both observations use J1517-2422 as the bandpass calibrator. Data reduction was done entirely using CASA software \citep{2008ASPC..394..623J}. Total flux accuracy at ALMA Band 3 usually better than 7$\%$. 
KK~Oph was calibrated using standard manual QA2 scripts, running in CASA 4.6.0, making use of the task fluxscale. The flux scale was determined from a compact QSO J1733-1304 which is monitored every 2 weeks.  At the time of observations it had a flux density 1.832~Jy, spectral index -0.65 at the reference frequency of '226.50259GHz.
HD~144667 was calibrated using the pipeline running in CASA 4.5.3. The flux scale was determined from Titan (Butler-JPL-Horizons 2012) using baselines shorter than 352 m.

As a test, we re-imaged HD~144668 using tclean. A marginally smaller synthesized beam (by 10~mas) was the only difference.   The phase reference was $\approx$1$^{\circ}$ from the target, PWV=2.52~mm. There was up to 20$^{\circ}$ change in phase reference solutions on individual antennas. The two 1.875~GHz spectral windows had enough S/N for phase self-calibration with a 60~s solution interval, i.e half the total time on target.  However, there was no improvement in S/N ($\approx$165 both before and after). For  KK~Oph,  the phase reference was $\approx$3~$^{\circ}$ separation, PWV=1.95~mm. The phase reference solutions showed a scatter within scans and a change between scans of less than 10$^{\circ}$ each, per antenna, averaging to less than 2$^{\circ}$ scatter in the corrected phases. The astrometric accuracy was determined as outlined  in Sect.~4 of \citet{2019MNRAS.485..739M}, and is 12-13 mas for both targets.
Residual phase errors after applying the phase reference source corrections affect the 'seeing'.  This smears the flux in proportion to the map noise, by approximately (beam size)/(signal-to-noise ratio), for snapshot observations such as these \citep{2017isra.book.....T,1997PhDT........20R}. Thus, for our resolution, the position uncertainties of emission at the 4 (3) sigma rms noise level are ~14 (18) mas.

Due to the complex structure of the brightness distribution, image reconstruction was performed interactively, using the \textit{clean} task in CASA Software Package. During clean, the Briggs robustness parameter (``\textsc{robust}'') was set to 0.5 and masks applied to enclose all detectable emission. Off-source rms achieved in the continuum images is 0.18~mJy/beam for HD~144668 dataset and 0.16~mJy/beam for KK~Oph. The detected emission structure is only a few arcsec across and as such does not require primary beam correction. CO isotopologue line emission was not detected in any of these observations, suggesting that any prior single dish detections may be due to cloud emission. Quantitative details on the upper limits follow in Sect.~\ref{gasmass}.

\begin{figure*}
    \hspace{-1cm}
    \includegraphics[width=2.1\columnwidth]{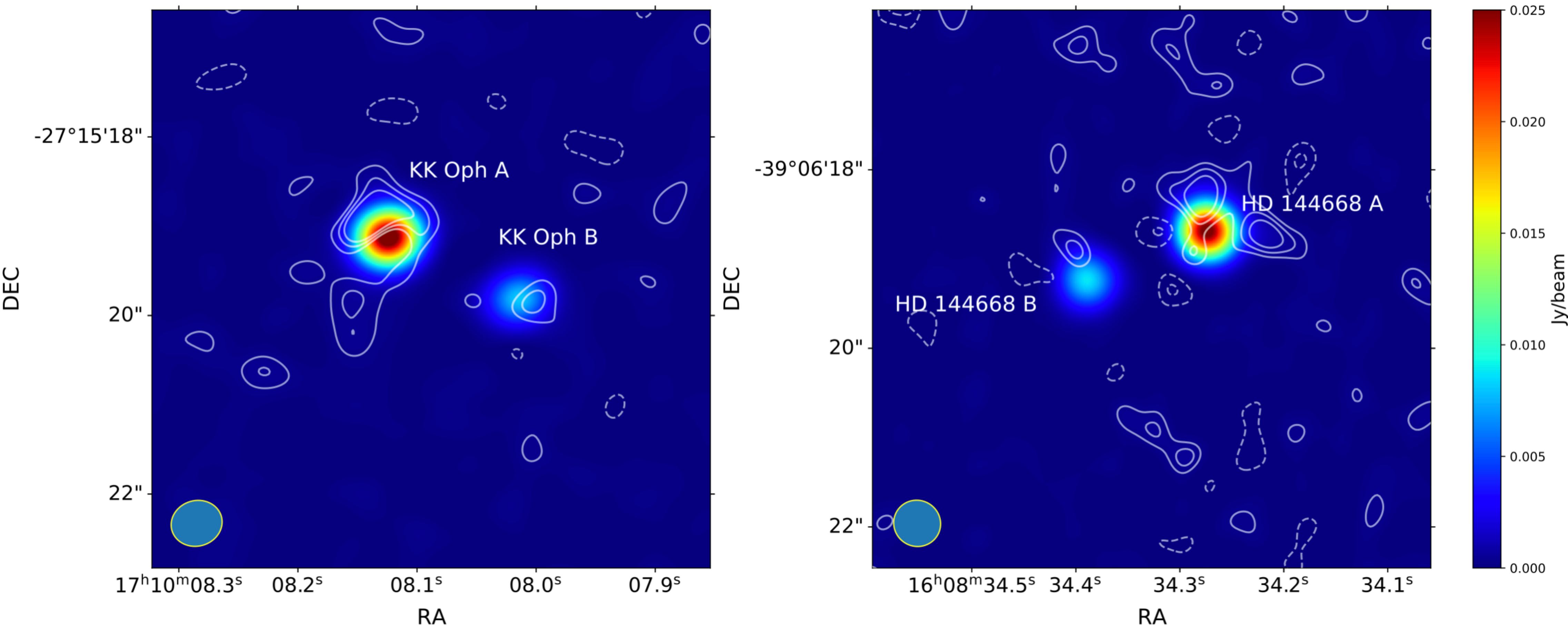}
    \caption{1.3~mm continuum images of the KK Oph (left) and HD 144668 (right) binaries is shown in colour. Overplotted in white contours are the residuals from the subtraction of the beam-convolved point source emission (see text). The contour levels for KK Oph are levels=(-3,-2,2,3,4)$\times$160$\mu$Jy/beam (1 sigma) and for HD 144668 the contour levels are levels=(-3,-2,2,3,4)$\times$180$\mu$Jy/beam (1 sigma). Dashed lines are used for the negative contours. The beam is illustrated by the solid colour ellipses in the lower left of each image. Note that the separation between the stars is 355~au (1$\farcs$5) in the KK~Oph system and 240~au (1$\farcs$5) in the HD~144668 system.  }
    \label{fig1}
\end{figure*}

Both the primaries HD~144668 and KK~Oph, and their companions were detected in 1.3~mm continuum (Fig.~\ref{fig1}). There is no detected emission that would firmly indicate a common envelope or a circum-binary structure in either of the systems. However such structure could not be excluded either as low-level emission, just below the detection threshold may well be bridging the short distance between the two discs. In fact, low-level emission at <2$\sigma$ connects the primaries with the secondaries but higher sensitivity observations would be needed to ascertain whether this is due to beam smearing or an underlying low-density common structure. Thus the detected 1.3~mm emission is clearly distinct for the primaries and secondaries in the current observations. In Tab.~\ref{tab1} the locations of the detected continuum peaks from our Band 6 data are shown, and fluxes listed.

In addition to our data we use archival Band 7 (0.87~mm, 330~GHz) continuum observations of HD~144668, which were taken as part of the ALMA Lupus survey \citep{Ansdell2016ALMAMasses,Miotello2016LupusDepletion}. These observations have an angular resolution of 0$\farcs$31x0$\farcs$28 and PA=-88$^{\circ}$. Continuum rms 0.53~mJy/beam for 4.8~GHz bandwidth. 

\subsection{Optical and near-infrared archival data}\label{opt}

To establish the precise locations of the stars in the observed systems, we use the publicly available data from the Gaia\footnote{This work has made use of data from the European Space Agency (ESA) mission
{\it Gaia} (\url{https://www.cosmos.esa.int/Gaia}), processed by the {\it Gaia}
Data Processing and Analysis Consortium (DPAC,
\url{https://www.cosmos.esa.int/web/Gaia/dpac/consortium}). Funding for the DPAC
has been provided by national institutions, in particular the institutions
participating in the {\it Gaia} Multilateral Agreement.} and ESO archives\footnote{Based on observations collected at the European Southern Observatory under ESO programme(s) 076.C-0708(A), 
and 095.C-0658(A).}. These are then compared to the position of the millimetre peaks in our ALMA data in Sect.~\ref{geometry}. While Gaia is used to obtain precise stellar positions of the primaries, NACO data are essential for determining the offset between the primary and secondary star, and the position angle of the system - values we will compare to the millimetre results, especially because reliable Gaia data are not available for all stars.

\subsubsection{KK~Oph system} 

KK Oph is an example of an astrometrically problematic source in the Gaia catalogue. It is not present in the Hipparcos catalogue and Gaia DR2 assigns to it a distance of 221.1$^{+12.4}_{-10.4}$~pc \citep{2018A&A...620A.128V}, though it includes the warning label that it might be unreliable.
\citet{Fairlamb2015ARates} derived a distance of 279$^{+86}_{-81}$~pc by placing KK~Oph on the ZAMS. This means that they excluded the lower range of stellar luminosities as inconsistent with the stellar evolution models. This distance is consistent with the aforementioned Gaia DR2 distance. Hereby we adopt the distance of 221~pc, with a note that this star has problematic stellar photometry and hence this distance estimate carries large uncertainties.

For the companion star, KK~Oph~B, there are no reliable measurements. Instead, we use high resolution adaptive optics imaging at 2~$\mu$m available in the ESO archive, obtained using the VLT/NACO instrument. For the KK~Oph system these images have a typical image quality of 75~mas spatial resolution. We determined the central positions (in detector coordinates) of KK~Oph~A
and KK~Oph~B on the pipeline-reduced NACO images using stellar profile fitting. The
precision of the position measurement is 0.2~pixels, which is 2.7~mas at
the given pixel scale.
We obtain a separation of 1$\farcs$6088$\pm$0$\farcs$0075, i.e., 356$\pm$2~au projected separation at the adopted distance of 221~pc. Additional uncertainties in the plate-scale, projected versus true separation, distance are not included in the $\pm$2~au error.

As we will see in the following Sections, the mm coordinates we obtain with ALMA are fully consistent with the above. 

\subsubsection{HD~144668 system}

HD~144668~A has a reliable astrometric solution in Gaia Data Release 2 (Lindegren et al. 2018), as we verified by computing the renormalised unit weight error (Gaia technical note Gaia-C3-TN-LU-LL-124-01). 
Its coordinates are R.A. = $16^{\rm h}08^{\rm m}34.275^{\rm s}$, Dec. = $-39^\circ06'18.683"$ for epoch=J2015.5. The errors on the R.A. and Dec are 0.064~mas and 0.022~mas respectively. The proper motions are: pmra=-8.95$\pm$0.15~mas/yr and pmdec=-23.007$\pm$0.076~mas/yr. 
Parallax of 6.207$\pm$0.070~mas gives the distance of 161.1$^{+3.1}_{-2.9}$~pc, which we will adopt throughout the paper for both HD~144668~A and HD~144668~B. This is consistent with the previous estimate of 160$\pm$32~pc \citep{Fairlamb2015ARates}.

Regarding the companion, HD~144668B, the renormalised unit weight error tells us that its Gaia astrometric solution is not reliable.

To assess the separation of the two stars in the near-infrared, we use the archival NACO data. 
Based on the profile fitting to the saturated PSF wings we obtain a separation of 1$\farcs$46$\pm$0$\farcs$02. Assuming the distance of 161.1~au the projected separation is 235$\pm$3~au.

\begin{table*}
 \centering
  \caption{Summary of the millimetre observations of KK~Oph and HD~144668. Peak and total integrated fluxes at 1.3~mm measured towards our targets, along with the best-fit Gaussian peak locations and sizes in comparison with the resolution of our observations. The slight differences in coordinates are within the astrometric precision of ALMA. Epoch corresponds to 2016.
* marks the data from \citet{Ansdell2016ALMAMasses}.}
  \begin{tabular}{@{}lccccccccccc@{}}
  \hline \hline
 Object & RA & Dec & $\lambda$ & Integrated flux & Peak flux  & Gaussian FWHM & Synthesized beam \\
  & (hh:mm:ss)  & (dd:mm:ss) & (mm) & (mJy) & (mJy/beam) & ('')  & ('') \\
\hline
HD~144668 & 16:08:34.274  & $-$39.06.18.68
& 1.2 & 26.25 & 25.61$\pm$0.18  & 0.532 x 0.525 & 0.53 x 0.52 \\
HD~144668$^{*}$ & 16:08:34.275 & $-$39:06:18.68
& 0.87 & 56.0$\pm$1.1 & 54.02$\pm$0.53  & 0.32 x 0.28 & 0.31 x 0.28 \\
HD~144668B & 16:08:34.390  & $-$39.06.19.22
& 1.2 & 9.27 & 8.74$\pm$0.18 & 0.545 x 0.530 & 0.53 x 0.52 \\
HD~144668B$^{*}$ & 16:08:34.391 & $-$39:06:19.23
& 0.87 & 16.1$\pm$1.0 & 16.35$\pm$0.53 & 0.31 x 0.28 & 0.31 x 0.28 \\
KK~Oph  & 17:10:08.124 & $-$27.15.19.12
& 1.2 & 28.90 & 27.03$\pm$0.16 & 0.606 x 0.598 & 0.58 x 0.51 \\
KK~Oph B & 17:10:08.014 & $-$27.15.19.80
& 1.2 & 8.3 & 8.1$\pm$0.16 &0.602 x 0.499 & 0.58 x 0.51 \\
   \hline
\end{tabular}
\label{tab1}
\end{table*}

\section{Results}\label{results}

\subsection{Geometry of the two binary systems}\label{geometry}
Millimetre emission in these two systems arises entirely from the discs and not the stars, so discrepancies with respect to the stellar position (Sect.~\ref{opt}) may arise as a result of asymmetries for example. The majority of young discs observed with a modest resolution, as is the case here, present a centrally peaked brightness distribution. We will use the coordinates of the millimetre peaks from Tab.~1, to assess the axial-symmetry of the millimetre disc emission in Sect. 4.2.

For KK~Oph~A, the millimetre emission peak reported in Tab.~1. is fully consistent with the stellar location from Gaia data, taking into account proper motion. Consistency is found also for KK~Oph~B where NACO data were used. From the Gaia and NACO data we obtain a binary separation measurement of 1$\farcs$6088$\pm$0$\farcs$0075. This measurement is consistent with the 1$\farcs$62 separation between the millimetre peaks.

Our mm-imaging does not allow us to derive the disc inclination. However it is worth noting that in the KK~Oph case the excess emission is located along the direction of the inclination of the sub-au sized disc around the star seen in optical and mid-infrared interferometry, which in turn is aligned with the direction of the companion \citep{2013A&A...551A..21K}. Our data are therefore consistent with the previous claims of high inclination of the disc around KK~Oph.

For HD~144668~A, we also have the millimetre peak at a location consistent with the stellar coordinates obtained from Gaia. The projected separation between the primary ansd secondary measured using NACO is 1$\farcs$46$\pm$0$\farcs$02 as discussed in Sect.~\ref{opt} and fully consistent with the separation between the two millimetre peaks in ALMA imaging of this system.

\subsection{Disc masses}

\subsubsection{Dust mass}
\label{sec:dustMass}

Using the 1.2~mm continuum emission fluxes $S_{1.2mm}$ (Tab.~\ref{tab1}) we calculate the dust mass of the discs using the equation relating dust mass and millimetre flux in the optically thin regime:
\begin{equation}
M_{dust}=S_{\nu}D^2/{\kappa _{\nu}}B_{\nu}(T)
	\label{eq1}
\end{equation}
where $\kappa _{\nu}$ is the dust opacity per gram of dust at frequency $\nu$ and $B_{\nu}$ is the Planck function at temperature $T$, for which we adopt $T=$20~K. Given the high S/N of these continuum detections (50-170, see Tab.~1)), the largest source of uncertainty in the dust mass estimates is the assumed dust opacity. As the study of \citet{2006ApJ...636.1114D} shows, the maximum opacity (corresponding to a millimetre-sized maximum grain size) is around 1.15~cm$^2$/g$_{dust}$, and we adopt this value in our calculations. We report the corresponding minimum dust mass estimates in Tab.~3, with a note that these values may be up to 50$\%$ higher due to uncertainties in the opacity.  
Considering that the dust opacity decreases below 1~cm$^2$/g$_{dust}$ for pebbles larger than 1~cm, larger amounts of solids may be present in these discs in the form of pebbles, rocks and larger objects. The millimetre emission measurements therefore serve as probes of \textit{dust} mass and not the \textit{solid} mass of the disc. 

\begin{table}
 \centering
  \caption{Gas to dust ratios, dust and gas masses derived in this work. Dust masses carry a 10$\%$ error due to flux calibration uncertainty and present a firm lower limit to the total amount of solids in the disc. Upper limits on the gas mass are based on our $^{13}$CO non detections, assuming a 20~K temperature. The gas to dust ratios are therefore the highest possible values given these observational constraints. 
    }
  \begin{tabular}{@{}lllllccccccc@{}}
  \hline \hline
  Object & M$_{dust}$ & M$_{gas}$ & g/d \\
 & (M$_{Earth}$) & (M$_{Jup}$) & \\
 \hline
 HD~144668 & 32.6 & $\le$0.087 & $\le$0.9\\
 HD~144668B & 11.4 & $\le$0.087 & $\le$2.4\\
 KK~Oph & 66.6 & $\le$0.045 & $\le$0.2\\
 KK~OphB & 19.2 & $\le$0.045 & $\le$0.8\\
\hline
   \hline
\end{tabular}
\label{tab3}
\end{table}

\subsubsection{Comparing millimetre spectral indices of the HD~144668 circum-primary and circum-secondary discs}\label{beta}

We measure the millimetre spectral index, which can be used to assess the degree of dust evolution \citep[see][for a full discussion on this]{2006ApJ...636.1114D}.  The flux $F_\nu$ scales with frequency $\nu$ according to the spectral index $\alpha$
\begin{equation}
        F_2 = F_1\left(\frac{\nu_2}{\nu_1}\right)^{\alpha}. 
        \label{equn:fluxSpec}
\end{equation}
Similarly the opacity $\kappa_\nu$ scales with frequency according to the spectral index $\beta$
\begin{equation}
        \kappa_2 = \kappa_1\left(\frac{\nu_2}{\nu_1}\right)^{\beta}. 
\end{equation}
We refer to $\alpha$ and $\beta$ as the flux and opacity spectral indices respectively. In the Rayleigh-Jeans regime, they are related by $\alpha = 2 + \beta$ \citep{2006ApJ...636.1114D}.

The opacity spectral index is also related to the power of the grain size distribution 
\begin{equation}
    \frac{dn(a)}{da}\propto a^{-p}
    \label{equn:grainDist}
\end{equation}
via
\begin{equation}
    \beta = \beta_{ISM}(p-3)
\end{equation}
where $\beta_{ISM}\approx1.7$ as discussed by \cite{2006ApJ...636.1114D} and $n(a)$ is the number of grains of size $a$. Note that in the Rayleigh-Jeans regime this is only valid for $3<p<4$ ($\beta>0$) and a $\beta < 0$ implies that the assumption of being in the Rayleigh-Jeans regime is incorrect. 

So with a measure of $\alpha$ we can compute $\beta$ which allows us to infer the power law of distribution $p$. For discs with growth to millimetre sizes $\beta$ is typically $\beta\sim 0.5$, corresponding to $p\sim3.3$ \citep[e.g.][]{2016ApJ...821L..16C}. These values are below the ISM ones, and indicate grain growth.

For the the HD~144668 primary $\alpha=2.35$, $\beta=0.35$ and $p=3.2$, values which are fairly typical of protoplanetary discs and are usually interpreted as being indicative of grain growth\footnote{Note that $\beta$ is sensitive to both the maximum grain size and the power law slope of the grain size distribution $p$, see Figure \ref{fig:opacities}} \citep{2014prpl.conf..339T}. For the the secondary (HD~144668B) with our assumption of the disc being the same temperature (20\,K) we find $\alpha=1.71$, $\beta=-0.29$ and $p=2.8$, which is intriguing because a) it is very different to the spectral indices of the primary and b) because it would be incompatible with being in the Rayleigh-Jeans regime. We hence next adopt a differential measurement of the spectral indices to negate flux calibration issues and measure the difference between, rather than absolute value, of the primary and secondary spectral indices.

\subsubsection{Differential measurement of $\beta$ between the circum-primary and circum-secondary discs}

 Derived spectral indices $\alpha$ and $\beta$, as well as the resulting power law of the grain size distribution $p$, suffer from two major sources of uncertainty: a) the absolute flux calibration of 10-15$\%$ in each measurement, exceeding the noise contribution seen that we have very high S/N in all detections, and b) the unknown temperature at which the emission arises in each disc. 

In this situation, a way to remove the calibration uncertainty is to use the ratio of flux of the primary to the flux of the secondary in each image to draw comparisons between the two discs, rather than measuring their individual $\alpha$ and $\beta$ values. We can therefore use the binarity to our advantage and calculate the difference in $\beta$ between the two sources using ratios of their millimetre fluxes. 

For observing frequencies $\nu_1$ and $\nu_2$ sufficiently close to assume the $\beta$ to be constant, and following from Eq.\ref{eq1}, we have:
\begin{equation}
\frac{\kappa_{\nu1}}{\kappa_{\nu2}}=\left(\frac{\nu_1}{\nu_2}\right)^{\beta}=\frac{S_{\nu1}}{S_{\nu2}}\frac{B_{\nu2}(T)}{B_{\nu1}(T)}\approx\frac{S_{\nu1}}{S_{\nu2}}\left(\frac{\nu_2}{\nu_1}\right)^2
\label{eq2}
\end{equation}
where we approximate $e^{-\frac{h\nu}{k_bT}}\approx 1-\frac{h\nu}{k_bT}$, valid for $\frac{h\nu}{k_bT}<<1$. Another assumption used above is that emission at both wavelengths arises from the same temperature (i.e., layer) within each of the discs. This is a valid assumption in case of optically thin millimetre emission, as emission at both 1.2 and 0.87~mm will be dominated by the bulk of the mass contained in the disc midplane. The parameter space in the luminosity-radius relation of \citet{2017ApJ...845...44T}, occupied by protoplanetary discs in the same millimetre luminosity regime as HD~144668 and HD~144668B all correspond to disc models with 0.87~mm optical depth lower than 1, supportive of our assumption that the emission is optically thin. 

Now, labeling the millimetre flux $S_{\nu}$ for the primary and secondary discs with $A$ and $B$ respectively, and frequency $\nu$ with 1 and 2 for 230 and 345~GHz (1.3 and 0.87~mm wavelength) respectively, we  find a simple relation between the flux ratios of the two discs at the two wavelengths and their millimetre opacities, $R$ as:
\begin{equation}
R=\frac{S_{\nu1}^AS_{\nu2}^B}{S_{\nu2}^AS_{\nu1}^B}=\left(\frac{\nu_1}{\nu_2}\right)^{\left(\beta^A-\beta^B\right)},
\label{eq3}
\end{equation}

We can now express the opacity difference between the circum-primary and circum-secondary disc as
\begin{equation}
\beta^A-\beta^B=\frac{\log R}{\log{\left({\nu_1}/{\nu_2}\right)}}.
\label{eq4}
\end{equation}
Applying the above reasoning to the HD~144668 binary system, for which we have measurements at two wavelengths, we obtain a very precise $\beta^A-\beta^B$=0.69$\pm$0.13. This is close to the difference we found in our direct estimate $\beta$ in section \ref{beta} and if correct could suggest that the dust is on average more evolved in the disc around the secondary star. Interestingly the magnitude of this difference for these coeval systems is similar to the typical range of spectral index measurements for T Tauri stars, as in Figure 5 of \cite{2014prpl.conf..339T}. This would suggest that the diversity in $\beta$ in unresolved samples isn't simply a function of age. 

If we assume the disc around the primary is optically thick at 0.87~mm - for example if only 37$\%$ of the emission escapes the disc (i.e., $\tau_{0.87mm}=$1), we adjust the Eq.~\ref{eq3}  such that 
\begin{equation}
R=0.37\times\frac{S_{\nu1}^AS_{\nu2}^B}{S_{\nu2}^AS_{\nu1}^B}=\left(\frac{\nu_1}{\nu_2}\right)^{\left(\beta^A-\beta^B\right)},
\label{eq3b}
\end{equation}
then $\beta^A-\beta^B$ difference becomes as large as 3.7, allowing for even more evolved dust grains in the disc around the secondary star in comparison to the disc around the primary. 

So, given a $\beta$ difference of 0.69 (accounting for flux calibration) we require either the primary to have a higher $\beta$ than is usually observed in discs ($>1$), or the secondary to have a lower $\beta$ and hence potentially more evolved dust. Alternatively the secondary may not be in the Rayleigh-Jeans regime at 0.87\,mm, in which case follow up observations are required to re-evaluate the secondary spectral index. 

If we compare the emission of these two stars to the millimetre size-luminosity relation found for protoplanetary discs \citep{2017ApJ...845...44T}, we would expect the secondary to have more than two times smaller disc than the primary, due to its lower millimetre flux. Having less material at large orbital distances, the secondary would then miss the more pristine reservoir of dust which in the case of the primary dominates the dust mass. This may be the explanation for, on average, flatter dust size distribution in the disc around the secondary. However, this raises the question of how such a compact disc can still retain a significant reservoir of large grains without them in-spiralling onto the star.

\begin{figure}
    \centering
    \includegraphics[width=9cm]{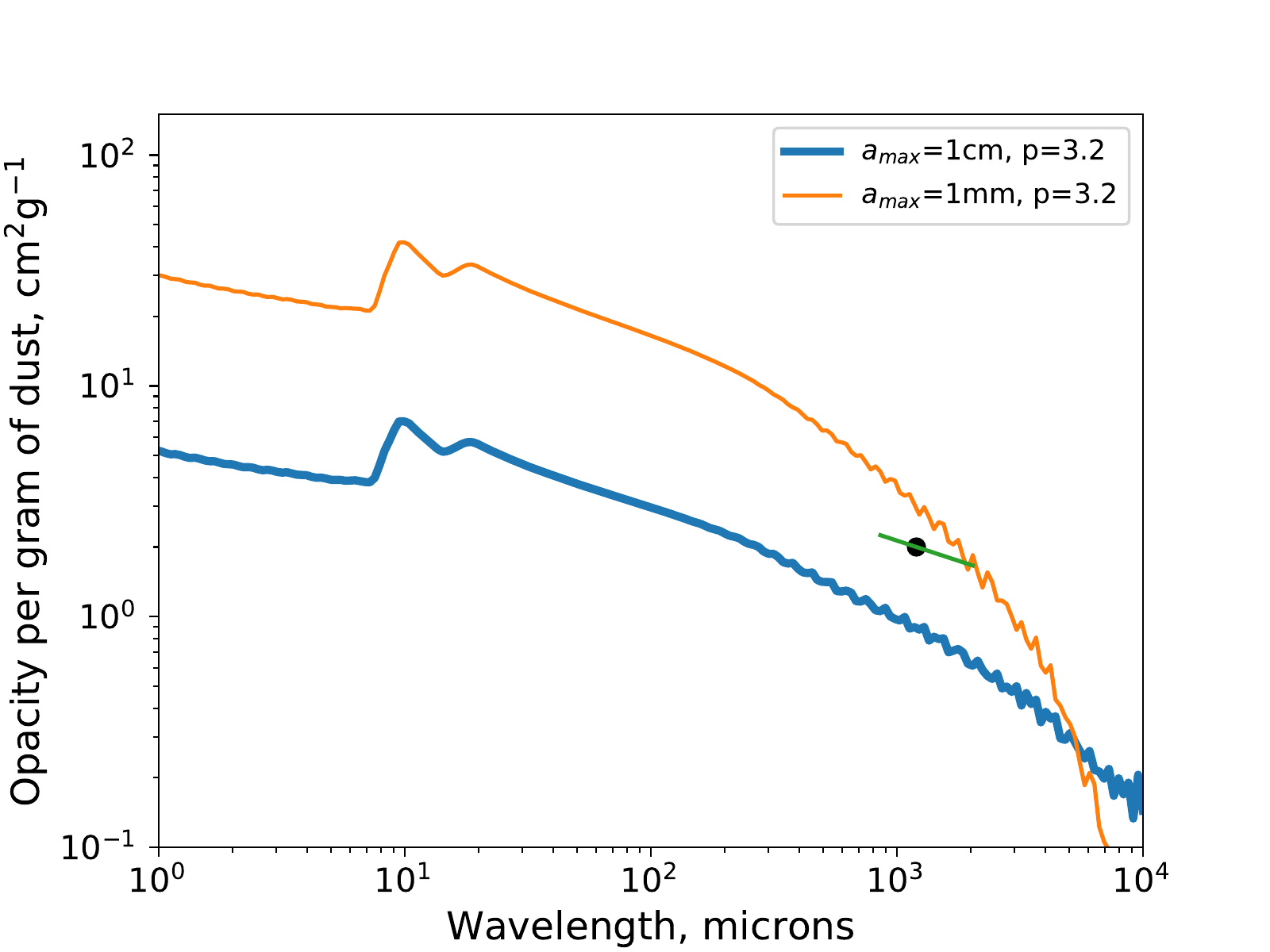}
    \includegraphics[width=9cm]{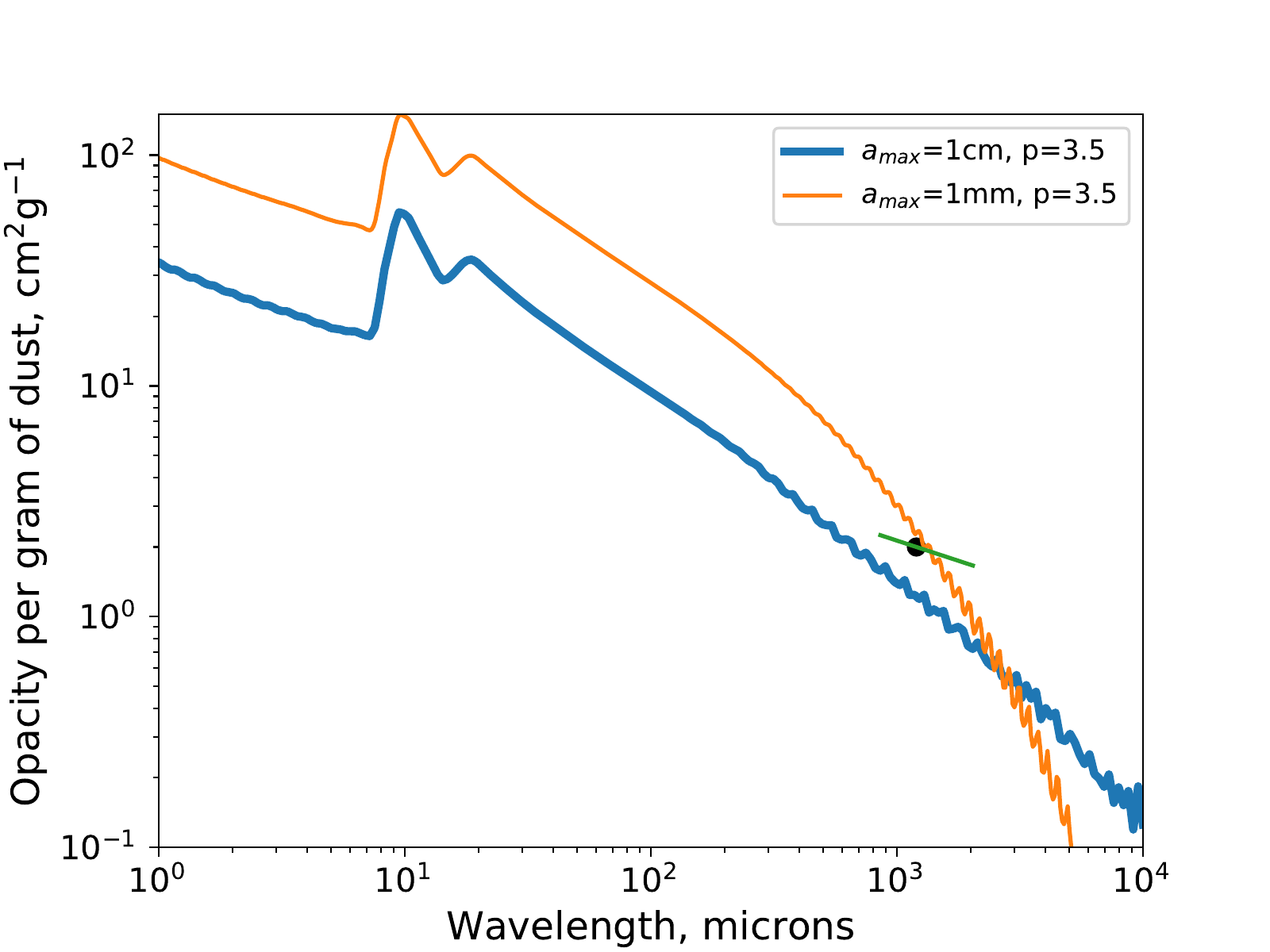}    
    \caption{The absorption opacity (per gram of dust) for the HD 144668 primary (upper panel) for amorphous silicates. The maximum grain sizes are $1\,$cm (blue) and $1\,$mm (orange). The opacity/wavelength assumed in our dust mass calculations is given by the black point and the line going through it has slope corresponding to $\beta=-0.35$, as we find for the HD 144668 primary. The lower panel is the same as \protect\cite{2006ApJ...636.1114D} which uses a different power law for the distribution.  }
    \label{fig:opacities}
\end{figure}

The inferred spectral indices/grain distribution also has consequences for the opacity, which in turn is a key source of uncertainty in the disc dust mass estimates. A large number of factors can affect the spectral index \citep[e.g.][]{2016A&A...586A.103W}, so to make a practical assessment within the scope of this paper we used the \textsc{torus} code \citep{2019A&C....27...63H} to compute the opacity for different grain distributions. We always assume amorphous silicates with a minimum grain size of 3.5\AA, following \citet[][]{2006ApJ...636.1114D}, consider the cases of maximum grain size 1mm and 1cm, and use a power-law index of the grain size distribution of $p=3.2$ inferred for the primary in section \ref{beta}. The resulting absorption opacities for the primary is given in Fig.~\ref{fig:opacities}, compared with the \citet[][]{2006ApJ...636.1114D} result that uses $p=3.5$. The millimetre opacity when the maximum grain size is $1\,$mm is always $\sim2$\,cm$^2$\,g$^{-1}$ regardless of whether we use $p=3.2$ or $p=3.5$, conversely for larger maximum grain sizes the opacity drops significantly. 

The main uncertainty in the opacity hence stems from the maximum grain size rather than the power of the distribution. For the plausible upper limits of 1mm and 1cm we find opacity variations of a factor 3 and 4.5 for the primary and secondary respectively. Given the apparently evolved state of the grain distributions, it is prudent to assume that the maximum grain size is large and our choice of opacity of $2\,$cm$^2$\,g$^{-1}$ for the dust mass calculation in section \ref{sec:dustMass} will not contribute to an overestimate of the dust mass.

\subsubsection{Upper limits on gas mass}\label{gasmass}

\cite{Hales2014AFORMATION} report integrated $^{12}$CO 3--2 line emission of 7.2 and 0.6~Kkm/s towards HD~144668 and KK~Oph with APEX observations with beam size encompassing the entire binary system, cautioning that both may be affected by cloud emission. In fact, each system is located in the region of a core seen in C$^{18}$O, at 4.1~km/s \citep{1999PASJ...51..895H} and 3.6~km/s \citep{1999PASJ...51..871O}, respectively. Neither CO isotopologue line was detected in our observations. This supports the interpretation of these prior single-dish (large beam) detections of $^{12}$CO as being due to cloud emission. However some gas is present in these discs as evidenced by the O[I] 63.2~$\mu$m emission detected by \cite{Meeus2012AstrophysicsDiscs}. Due to the same rms levels for the measurements of the primary and the secondary (as they are both in the same image), the integrated emission over a wide range from 0 to 8~km/s is 14\,mJy\,km/s for $^{13}$CO J$=$2--1, for each star in the KK~Oph system and 51\,mJy\,km/s for $^{13}$CO J$=$2--1, for each star in the HD~144668 system.

Assuming local thermal equilibrium and optically thin emission we derive upper limits on the gas masses. It is a standard assumption that in protoplanetary discs $^{13}$CO low-J rotational lines are thermalised because of their optical depth, allowing them to trace the denser interior regions. In the case of these two discs, we may be in a situation of non equilibrium if the gas mass is as low as to allow the gas to become optically thin to stellar emission. In such case the approach developed for debris discs in \citet{2015MNRAS.447.3936M} would be more appropriate. The lowest temperature at which CO is expected to generate emission in this line in protoplanetary discs is the temperature at which CO freeze-out becomes efficient in removing CO from the gas-phase, around 20~K. CO gas in protoplanetary discs has been measured down to very low temperatures, i.e., $T<20\,$K \citep{2003A&A...399..773D, 2007A&A...467..163P}, while the warmest midplane temperature at the outer radius derived in physical modelling of discs reaches 25K at 150~au and as high as 50~K at 10~au for exceptionally small 10~au discs. We therefore adopt 20~K in our gas mass estimates, as this yields a firm upper limit, while any higher temperature would correspond to a lower mass estimate.

We estimate the gas mass as follows. The line integrated emission coefficient for a molecular transition from level $u \rightarrow l$ is
\begin{equation}
	j = \frac{h\nu_{ul}}{4\pi} A_{ul}n_u \hspace{0.5cm} (\textrm{e.g.} \hspace{0.1cm} \textrm{erg}\,\textrm{s}^{-1}\,\textrm{cm}^{-3}\,\textrm{st}^{-1})
\end{equation}
where $\nu_{ul}$, $A_{ul}$ and $n_u$ are the frequency of transition, Einstein A coefficient and number density of species in the upper excited state respectively\footnote{Molecular data from \citet{1998JQSRT..60..883P,2001A&A...370L..49M,2005A&A...432..369S}}. 

If the total mass of the emitting molecules  is $M_{\textrm{mol}}$ and the mass per molecule is $m_{mol}$ and the medium is optically thin, the flux at distance $d$ is simply
\begin{equation}
	F_\nu = {h\nu_{ul}} A_{ul}  \frac{x_uM_{\textrm{mol}}}{4\pi d^2 m_{\textrm{mol}}} \hspace{0.5cm}(\textrm{e.g.}  \hspace{0.1cm}  \textrm{erg}\,\textrm{s}^{-1}\,\textrm{cm}^{-2})
\end{equation}
where $x_u$ is the fraction of the molecule in state $u$, i.e. ($n_u/n$) and $m_{\textrm{mol}}$ is the mass of a single molecule. The detected flux $F_{ul}$ is hence related to the mass of the molecule in the object by
\begin{equation}
	M_{\textrm{mol}} = \frac{4\pi m_{\textrm{mol}} d^2 F_{ul}}{h A_{ul} x_u \nu_{ul}}.
	\label{eqn:quickMass}
\end{equation}
which can be converted to a total mass using the fractional abundance. The only inputs to this expression are the detected flux and $x_u$. We solve for the latter analytically using a Boltzmann distribution, assuming local thermodynamic equilibrium and computing the partition function up to level J=40, which is sufficiently high that less than one part per trillion are in this state at the adopted range of gas temperatures. We also assume optically thin emission in this estimate.

\begin{figure}
	\includegraphics[width=\columnwidth]{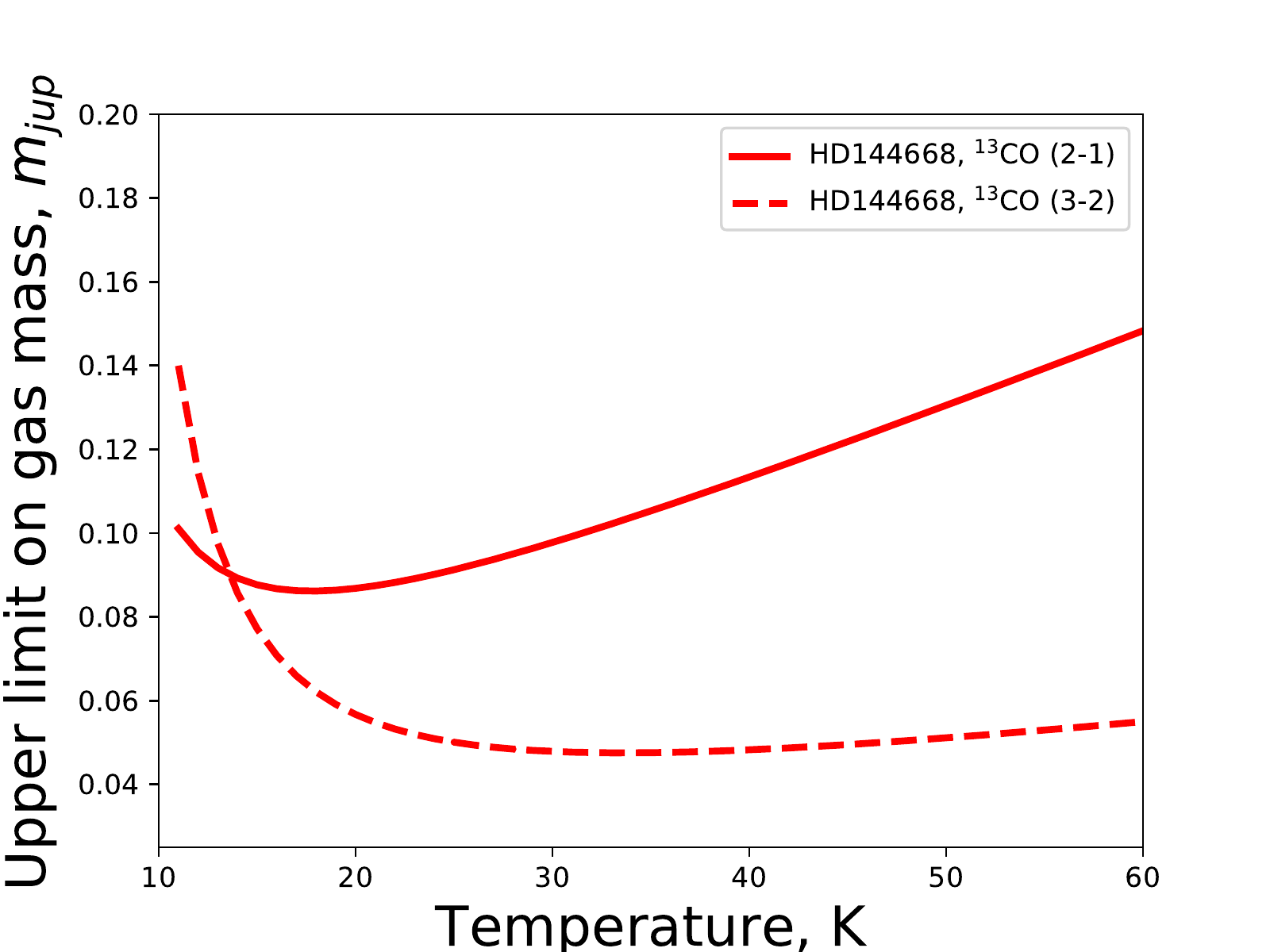}
	\includegraphics[width=\columnwidth]{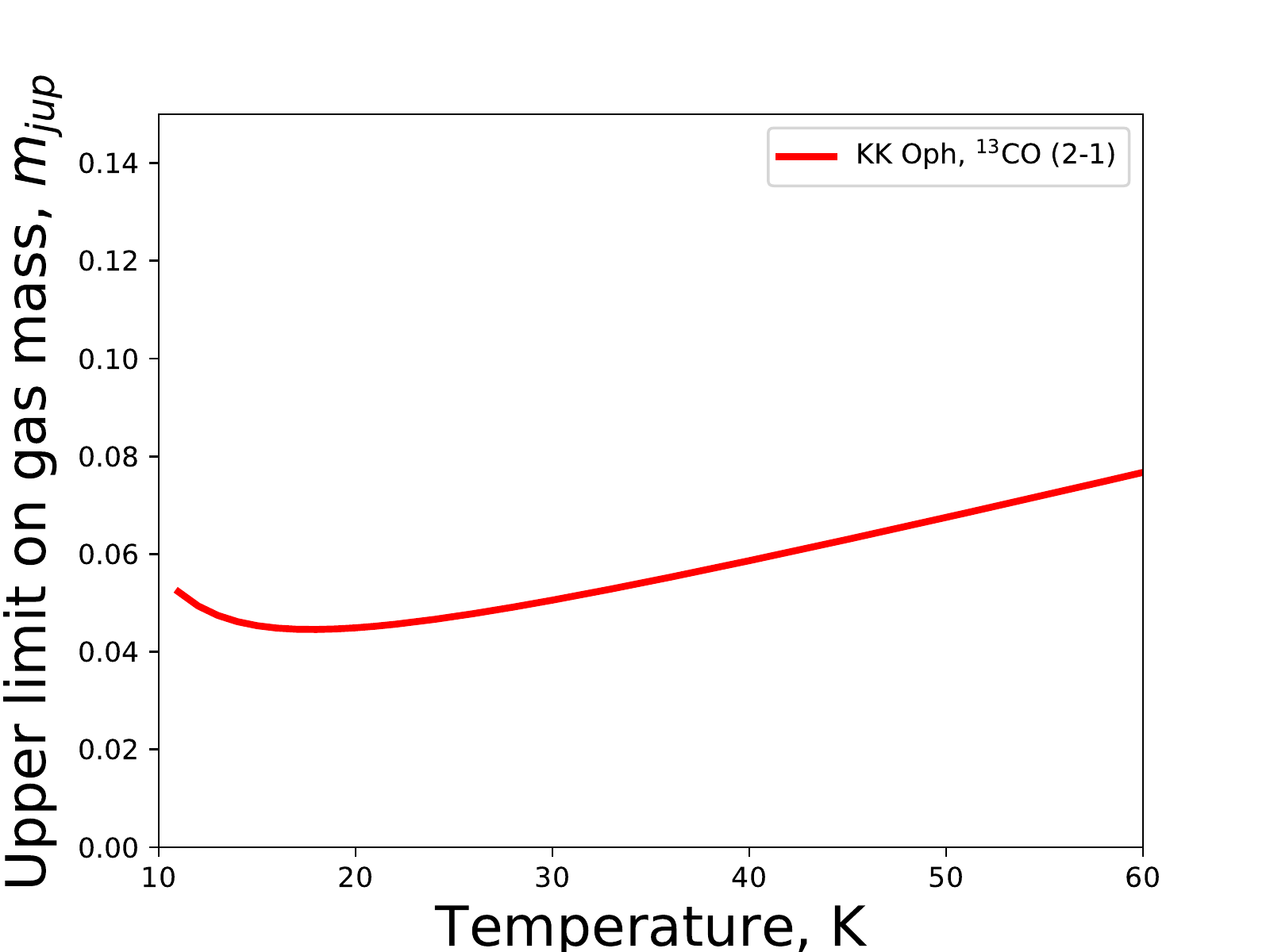}
\caption{Upper limits on the gas mass as a function of temperature for the individual discs from $^{13}$CO 2-1 (solid) and 3-2 (dashed) transitions. The upper panel is for HD\,144668 and the lower panel for KK Oph. For the latter system we do not have 3-2 data and the upper limit on the 2-1 flux is the same for the primary and secondary, so their mass estimates are coincident. For the estimates in Table \ref{tab3} we use the mass at 20\,K.  }
    \label{fig3}
\end{figure}

Using the described method we obtain firm upper limits on the mass of $^{13}$CO gas that is present in these discs. We convert from a $^{13}$CO mass to total mass using a relative abundance of $1.3\times10^{-6}$ \citep{1994ARA&A..32..191W}. Fig.~\ref{fig3} shows the upper limits on the gas masses for the adopted range of temperatures and Tab.~\ref{tab3} lists the minimum upper limits on the gas mass, $\leq$0.045~M$_{jup}$ for each star in the KK~Oph system and $\leq$0.087~M$_{jup}$ for each star in the HD~144668 system, as also shown in Tab.~\ref{tab3}, and obtained under the assumption of a 20~K temperature.

We note that for KK~Oph even the lower measured accretion rate of 
1.2 $\times$ 10$^{-8}$ M$_\odot$~yr$^{-1}$, respectively \citep{2006A&A...459..837G} would not be possible to sustain with this upper limit on the disc mass for more than about 35\,kyr. This implies that either the accretion is highly variable, or could be periodically driven by the binary interaction. We also note that these accretion rate measures do require a gas mass reservoir. 

The assumption of standard CO abundance and isotopic ratio may be underestimating the derived masses by a factor of at most 3-4, in case CO may largely be frozen out. This estimate comes from the mass contained in a disc beyond the CO snowline.  Another source of uncertainty is photodissociation. Using \textsc{dali}, \cite{2014A&A...572A..96M} (their Fig.~6) computed the ratio of line intensities of CO isotopologues with and without photodissociation in Herbig discs. They found that although C$^{17}$O and C$^{18}$O line ratios are strongly sensitive to the effects of photodissociation, $^{13}$CO is relatively unaffected out to 100\,au (our discs are less extended than this). We therefore have the theoretical expectation that photodissociation does not significantly affect our upper limits on the $^{13}$CO mass. 

To explore the scenario of foreground cloud absorbing a part of the emission from the disc and therefore causing the non-detection we investigate the line-of-sight extinction towards the two binary systems considered here.  According to the extinction maps of \citet{2006A&A...454..781L} and \citet{2008A&A...489..143L}, KK~Oph is not in a region of significant extinction, but there is some line-of sight extinction towards HD~144668. Using the dust map of \citet{2019A&A...625A.135L} and the Gaia DR2 distances and coordinates to the primary stars, we obtain an E(B-V)=0.046${\pm}$0.039 for KK~Oph and E(B-V)=0.015$^{+0.022}_{-0.015}$ for HD~144668 system. We adopt the maximum values possible for E(B-V) for each of the systems and use it to estimate the amount of material along the line of sight. As $A_v=3.1\times E(B-V)$ \citep{1975A&A....43..133S} and $N_{H}=2.2\times 10^{21}cm^{-2}A_v$ \citep{2009MNRAS.400.2050G} and adopting the ISM CO abundance of 10$^{-4}$ and isotopic ratio $^{13}$C/$^{12}$C$=$1:66 \citep{1989ApJ...344..311F} we obtain the line-of-sight column of $^{13}$CO$\approx10^{15}cm^{-2}$. The optical depth we obtain for the $^{13}$CO $J=$2--1 line for this column, and assuming local $H_2$ density of 10$^2$cm$^{-2}$ and a temperature of 20~K, is $\tau=$0.35 for HD~144668 and an order of magnitude lower $\tau$ . Considering that we used the maximum E(B-V), this is the maximum optical depth expected so we can safely assume that no line of sight absorption is affecting the emission from either HD~144668 or KK~Oph.

\subsection{Resolved asymmetries in the outer regions of the discs}\label{asymmetry}

Our observations have clearly resolved the emission of the circum-primary and the circum-secondary discs from one another. Visual inspection does not indicate that disc structure is significantly resolved, but there are extended areas of low-level emission in both systems which prompt us to do a more thorough examination. To establish the significance of the emission in these extended regions we subtract the emission expected from a point source, convolved with the synthesized beam. We first use CASA task \textit{imfit} to find the best-fit 2D Gaussian distribution in the region which isolates the emission of a disc from the emission of the disc of its companion. Gaussians obtained are almost circular, with FWHM of the major and minor axes listed in Table \ref{tab1}, which, for all discs closely resemble the synthesized beam size. We then proceed by placing point sources of appropriate flux at the best-fit locations of the Gaussians (also listed in Table \ref{tab1}) in blank images and convolve these images with the appropriate synthesized beam sizes. This is then subtracted from the ALMA images. For all observed stars the millimetre peaks are entirely consistent with the known optical coordinates of the corresponding pre-main sequence star as computed based on J2000 coordinates and proper motions provided in the SIMBAD database \citep{Wenger2000TheDatabase}, and reconfirmed by Gaia.

Subtraction of beam-convolved point source emission for the KK~Oph system also reveals excess emission up to 4~$\sigma$ level in the North-East direction of the primary (see Figure \ref{fig1}). Optical interferometry of KK~Oph also reveals a very inclined disc with the major axis nearly along the North-East South-West direction \cite{2013A&A...551A..21K}, which is also the direction towards the companion. 

In our Band 6 data for HD~144668 this procedure reveals a significant extent of emission stretching from the North to the West of the primary star reaching 4$\sigma$ levels (see Fig.~\ref{fig1}). This indicates a large-scale asymmetry in the outermost regions of the circum-primary disc, and a disc size that is very close to the spatial resolution of our observations, corresponding to the disc radius of 42~au at the distance of 161~pc. There is also a tentative detection of an excess North-East from the secondary just reaching 3$\sigma$, but the size of this excess is very small. Asymmetries such as these are expected in the outer reaches of the disc where the gravitational influence of the companion becomes non-negligible. We discuss the implications of these features further in Sect.~\ref{asymm2}. Similar procedure was applied to the Band 7 archival data, but no excess is seen. This is unsurprising, considering the lower S/N in the Lupus snapshot survey.

\section{Discussion}

\subsection{Dust mass}
While the dust masses of the circum-primary discs KK~Oph and HD~144668 are well above the debris disc regime, the discs around the secondaries have barely enough dust mass to form minimum mass cores for giant planet formation via core-accretion \citep{2006ApJ...648..666R}. In fact their dust masses are at the low end of the derived heavy element mass of giant exoplanets \citep{2016ApJ...831...64T}, which is not dissimilar from protoplanetary discs in general \citep{2018A&A...618L...3M}. What is striking about these systems is the lack of gas, as discs as dust-rich as these typically have a detectable gas reservoir, of around 0.01~M$_{\odot}$.

\subsection{Maximum gas mass from $^{13}$CO upper limits} 

Our gas mass limits are all below 0.1~M$_{jup}$ \textbf{(Tab.~\ref{tab3},} showing that giant planet formation is no longer possible in the two observed binary systems. The primary stars, HD~144668 and KK~Oph have the least massive gas discs in our larger survey of 15 intermediate-mass stars (Pani\'c et al in prep.), and are also the only intermediate separation binary systems in the sample. A meaningful comparison is that to the most evolved disc in our survey, HD~141569. This disc has a considerably higher gas mass than the upper limits for our binaries, 0.6~M$_{jup}$ of gas, and is characterised as a debris disc by its dust mass, SED and ringlike spatial structure in dust \citep{2018A&A...615L..10M}. One could view the gas-rich and dust-poor HD~141569 as the opposite result of disc evolution than our binaries, which we find to be gas poor and dust-rich in comparison.

The lack of gas in HD~144668 is unusual when compared to the other discs in Lupus. Our results imply a low gas to dust ratio of $g/d\leq$0.9. This upper limit on the gas-to-dust mass ratio is currently the lowest for a disc in Lupus, when compared to the findings for other observed Lupus discs in the \cite{Ansdell2016ALMAMasses} survey. There, 26 discs out of 62 continuum detections were not detected in any CO isotopologue lines, including HD~144668. Interestingly, the dust mass of HD~144668 is the highest amongst all Lupus non detections of CO. 

It is tempting to speculate whether the binary nature of our two systems led to the premature removal of gas with respect to the dust. One way in which this may happen is through dynamical interaction, discussed in further sections. Another is efficient mutual photoevaporation, for example if the two discs are mis-aligned allowing the radiation from the companion star to illuminate and externally photoevaporate the disc from great proximity. A literature search of intermediate-separation binary systems did not turn up any discs with gas masses as high as a few M$_{jup}$, though note that for many systems only dust mass estimates are available  \citep[e.g.,][]{2012ApJ...751..115H}.

\subsection{mm spectral index from B6/B7 for HD144668}
The range of the opacity spectral index $\beta$ expected in protoplanetary discs at millimetre wavelengths is from around 0.8 for larger, cm-sized grains, to 3.0 for 100~$\mu$m grains. The low value obtained for HD~144668~B in Sect~\ref{beta}, $-$0.29, may be indicative of a lower value of the power of grain size distribution $p$ (see equation \ref{equn:grainDist}), which allows for lower values of $\beta$ \citep{2006ApJ...636.1114D}. 

If we assume that the maximum grain size is comparable or larger than 1~mm, the large difference in $\beta$ between the primary and secondary is possible if the primary disc has $a_{max}=$1~mm and the secondary $a_{max}\ge$10~cm, which yields $\Delta\beta=$0.4, with $\beta^A=$1.2 and $\beta^B=$0.8 \citep{2006ApJ...636.1114D}. An alternative explanation would be that both discs have very evolved grains but that the disc with a lower value of $\beta$ has a flatter grain size distribution. \cite{2006ApJ...636.1114D} shows an example of astrosilicate with $a_{max}$=100~cm and $p=$3.0 and 3.5 which produce spectral indices of 0.2 and 0.8 respectively.

An interesting comparison can be made between the 3~Myr old HD~144668 binary system and the 2~Myr old 253-1536 binary in Orion. Both are intermediate-separation binary systems with an intermediate-mass primary and low-mass companion star. 253-1536 has a projected separation of 460~au, somewhat larger than the 235~au for HD~144668. Using the same approach as for the calculation of $\Delta\beta$ for HD~144688 in Sect.~\ref{results}, and fluxes of the 253-1536 A and B from \cite{2011ApJ...739L...8R} at 0.88 and 6.9~mm we obtain $\Delta\beta_{253-1536}$=0.04$^{+0.10}_{-0.05}$. Using the same 6.9~mm fluxes combined with newer 0.856~mm observations of \cite{2009ApJ...699L..55M} we obtain $\Delta\beta_{253-1536}$=-0.09$^{+0.11}_{-0.09}$. Combining these two constraints results in a very narrow range of $\Delta\beta_{253-1536}$=-0.05-0.02, which indicates that the grain growth in the primary and secondary is almost the same. This result for 253-1536 is in contrast with our results for HD~144668, where $\Delta\beta$=0.69$\pm$0.13 implies a significant difference between the two discs, with the primary having some reservoir of less evolved grains. Further observations and dedicated modelling will be required to determine the cause of this, but it may arise due to the secondary being a factor two more compact, which would correspond to more rapid dynamical processing of the dust.

\subsection{Tidal truncation of the primary disc}
\label{sec:truncationRadius}

Tidal truncation of discs in binaries can have a significant effect on their radial extent, which may affect the subsequent evolution and also the detected mass. We follow \cite{Harris2012ASystems}, \cite{Pichardo2005CircumstellarBinaries} to estimate the truncation radius of the primary via 
\begin{equation}
	R_t \approx 0.337 \left[ \frac{\left(1-e\right)^{1.2} \psi^{2/3}  \mu^{0.07}}{0.6\psi^{2/3} + \textrm{ln}\left(1+\psi^{1/3}\right)} \right] \mathcal{F} a_p
    \label{equn:truncationRadius}
\end{equation}
where $e$ is the eccentricity, $\psi$ is the mass ratio of the primary star to companion ($M_p/M_s$), $\mu$ is the reduced mass of the stellar pair ($\mu = \left\{M_s/M_p\right\}/\left[1 + \left\{M_s/M_p\right\}\right]$) and $a_p$ is the projected separation as viewed in the plane of the sky. $\mathcal{F}$ is the ratio of the true semimajor axis $a$ to the projected separation $a_p$, which \cite{Torres1999SubstellarApproach} show is
\begin{equation}
	\mathcal{F} = \frac{a}{a_p} = \frac{1}{1-e\cos(E)} \frac{1}{\sqrt{1-\sin^2\left(\omega+\nu_T\right)\sin^2 i}}
\end{equation}
where $\omega$ is the longitude of periastron, $E$ is eccentric anomaly, $\nu_T$ is true anomaly and $i$ the orbital inclination. For those unfamiliar, these (predominantly angular) parameters all just define the current position on an elliptic Keplerian orbit and are discussed further below. Note that the above equations can be applied to compute the truncation of the secondary disc by inverting the stellar mass ratio.

Since the orbital parameters in the above equations are unknown we follow \cite{Torres1999SubstellarApproach} and  \cite{Harris2012ASystems} and sample them using a Monte Carlo approach. That is, we construct an array of bins of possible disc radius (i.e. truncation radius) $R_t$. We then choose a set of random orbital parameters from which we compute a $\mathcal{F}$ and hence $R_t$, the latter of which is then added to the appropriate bin in the array. Doing so a large number of times and normalizing the resulting distribution across bins gives a probability distribution for the truncation radius $P(R_t)$. We use 1000 bins from 0 to 400\,au and 10 million sets of random orbital parameters to populate the distribution. 

The sampling of the orbital parameters is as follows.  $\omega$ is sampled randomly with uniform distribution from 0 to $2\,\pi$. We sample the eccentricity uniformly from 0 to 0.7, that is assuming that the orbit is not highly eccentric \citep{2006ApJ...646..523R}. The inclination has a sine dependency, hence a given inclination for random variable $r$ is $\textrm{asin}(r)$.

The \textit{mean} anomaly $M$ is constant with orbital phase and so is randomly sampled with uniform probability from 0 to $2\pi$. This is related to the eccentric anomaly $E$ by
\begin{equation}
	M = E - e\sin(E)
\end{equation}
from which we solve for $E$ by bisection starting from $E=\pi$. The true anomaly is then
\begin{equation}
	\nu_T = 2\, \textrm{atan}\left[\sqrt{\frac{1+e}{1-e}}\tan\left(\frac{E}{2}\right)\right].
\end{equation}
The ``known'' inputs required for these calculations are the projected separation and the stellar masses. We use the separations discussed earlier in the text: namely $1.6088\arcsec$ and $1.46\arcsec$ for KK Oph and HD\,144668 respectively with distances of 221 and 161\,pc. The stellar mass ratios are in the range 1.5--1.67 and 2.2--4.285 for KK Oph and  HD\,144668  respectively. We re-run the entire Monte Carlo sampling over the range of expected stellar mass ratios. 

There is a further constraint on our models, in that we know the primary discs are marginally resolved in both cases, so their projected sizes are larger than the corresponding beam sizes, which sets the lower limit on the disc size as 
72~au for KK~Oph and 
43~au for HD\,144668. Similarly, since the secondaries are point sources, the upper limit on the secondaries are 72\,au for KK~Oph and 43\,au for HD\,144668. 

\begin{figure}	
\includegraphics[width=9.4cm]{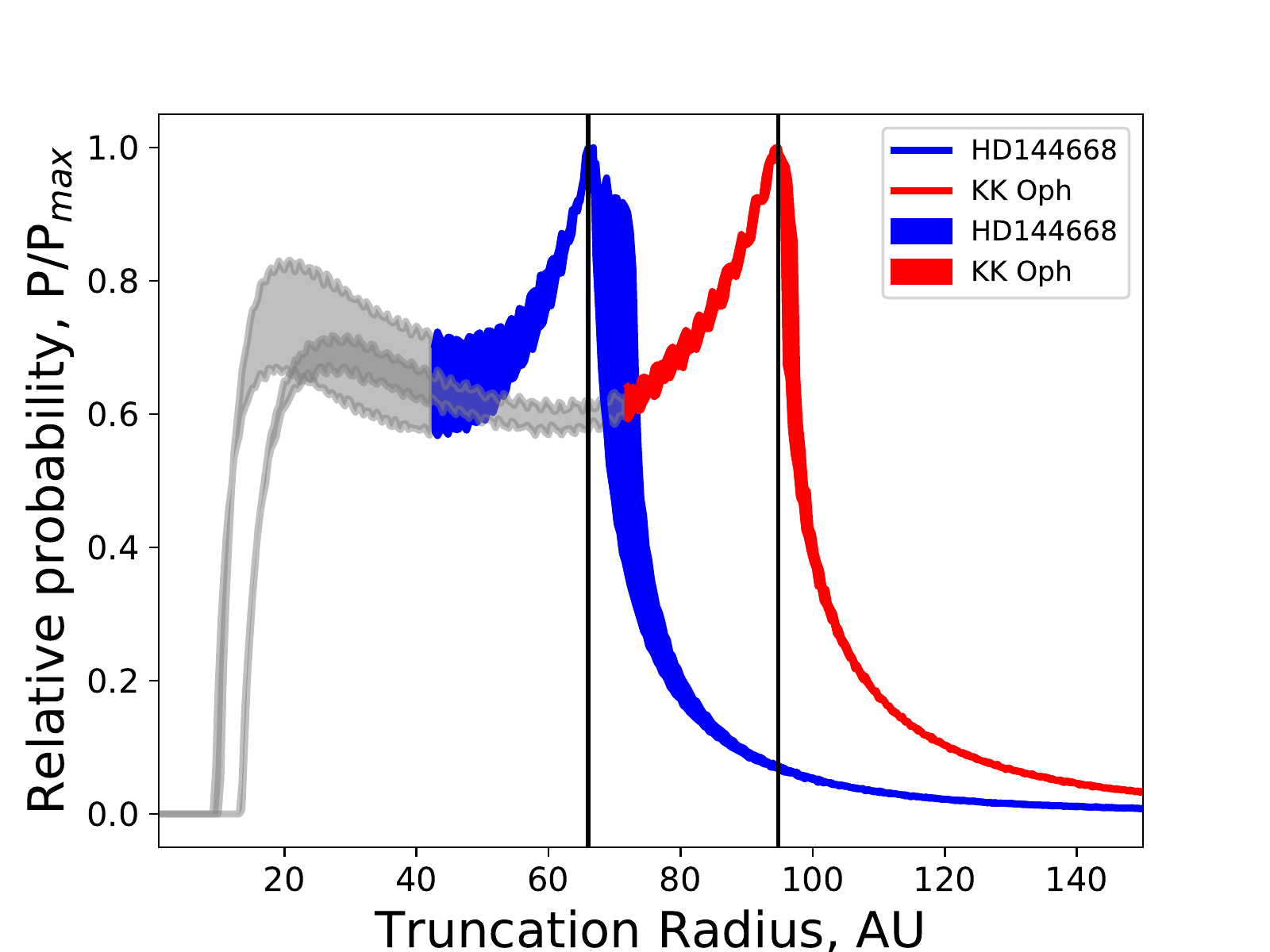}
\includegraphics[width=9.4cm]{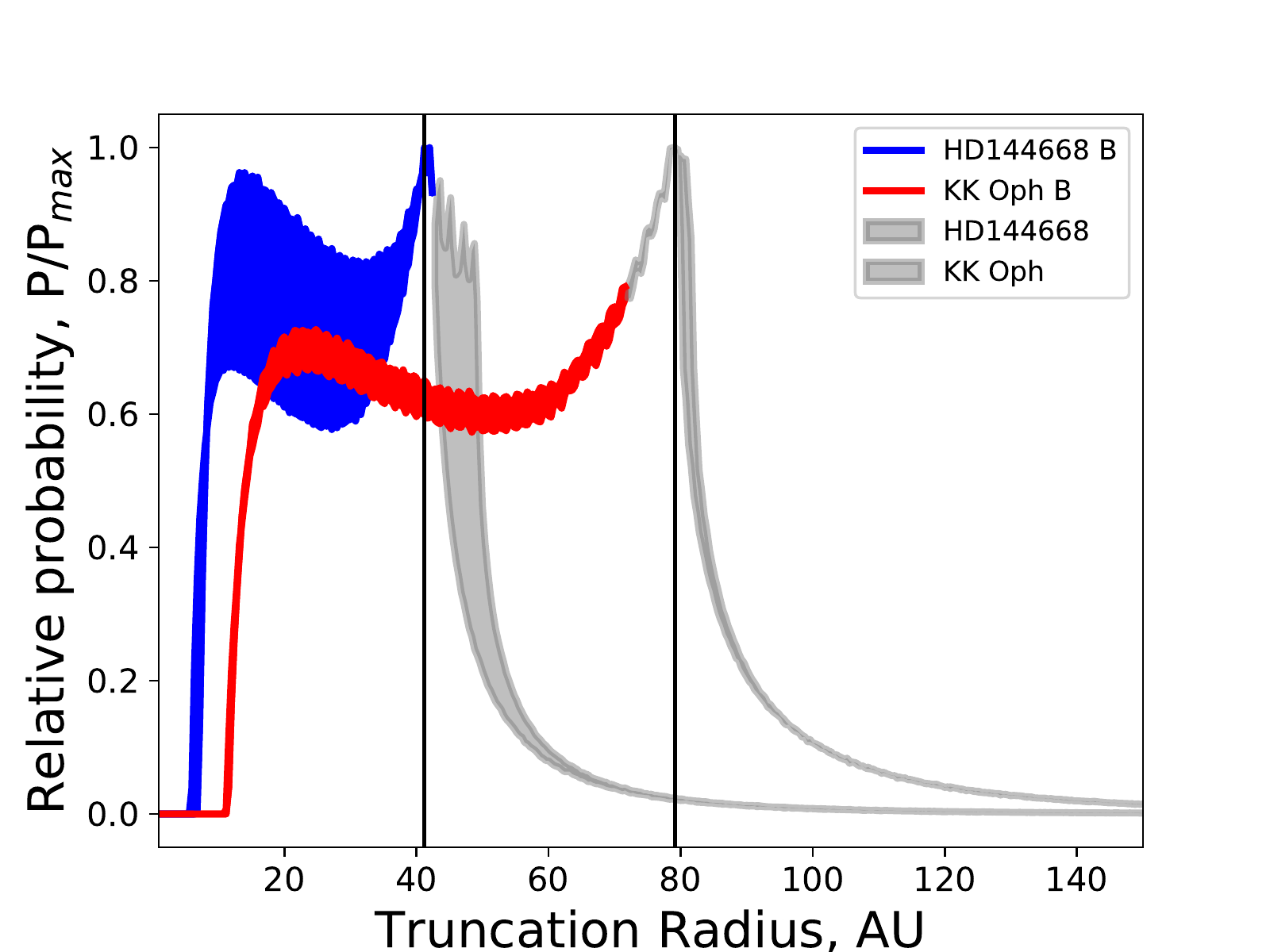}
  \caption{Normalised probability distributions of the truncation radius of the primary (upper panel) and secondary (lower panel) discs of HD 144668 (blue) and KK Oph (red). The primaries are resolved, but the secondaries are not, which allows us to place constraints on the disc size relative to that of the beam. This places lower limits on the primary disc sizes (denoted by the gray shaded region, which goes up to the beam size) and upper limits on the secondary disc sizes (again, denoted by the gray region in the lower panel, which is beyond the beam size).  The vertical lines denote the peak probabilities. }
  \label{RtProbability}
\end{figure}

The resulting probability distributions in the truncation radii of the primary and secondary discs for KK~Oph and HD\,144668 are shown in Figure \ref{RtProbability}. In both cases the region excluded due to whether or not the disc is resolved by the beam is coloured gray. The most probable truncation radii (and hence actual disc radius) of the primaries are 95 and 66\,au respectively. It is immediately clear that the primary disc of HD\,144668 is expected to be more truncated than that of KK\,Oph. This might explain the initially surprising result that the younger, HD~144668 system has the lower dust mass. Stronger truncation will result in faster growth and radial drift of grains \citep{2012A&A...539A.148B}. This could lead to faster growth of grains to a size at which solids are not detected at millimetre wavelengths. Additionally, if there is no pressure bump to trap drifting grains they could also end up in the very inner disc where collisions fragment them \citep[see e.g. Figure 3 of][]{2019MNRAS.484.2296J} rendering them undetectable at mm wavelengths and possibly end up accreted onto the star.

\subsection{Asymmetry of the primary disc}\label{asymm2}
In Sect.~\ref{asymmetry} we find extended features in the dust continuum emission in the circum-primary discs in both KK~Oph and HD~144668, as shown in Fig.~\ref{fig1}. In both cases the extended emission is located away from the projected direction of the binary companion, which is consistent with hydrodynamical models of dynamical perturbation of protoplanetary discs in fly-by scenarios \citep{2015MNRAS.449.1996D, 2019MNRAS.483.4114C, 2019MNRAS.tmp.2545C}.
In the case of our discs, the maximal perturbation induced by the companions on the outermost regions of the circumprimary discs would be achieved if the projected separation is taken as the orbital distance. The comparison between the angular resolution and projected distance in Tab.~\ref{tab1} implies that the ratio between the orbital period in the outer disc (adopting beam size as the maximum size of the disc) and the orbital period of the binary are around 1.2 for both systems. Therefore the outer disc does not have enough time to circularise following the encounters with the companion, whereby, for example the companion has an elliptical orbit such to periodically perturb the circumprimary disc. An example of a similar morphology seen in observation is in the case of the Sr24 binary \citep{2010Sci...327..306M}
where some extended scattered light emission is seen in the primary disc, pointing away from the companion. Such flyby scenarios also predict the discs to be warped which would help reconcile the high line of sight extinction towards KK~Oph with the outer disc morphology we see here, which is far from being edge-on. Further, higher resolution imaging of these systems with ALMA will be key to disentangling the exact morphology and sizes of their discs, explaining the dynamical processes which shape them. 

\section{Summary}
We present the first ALMA observations of the intermediate separation Herbig binary systems KK~Oph and HD~144668 to spatially resolve the discs (circum-primary and circum-secondary) from one another. These intermediate separation binaries, with each star hosting its own protoplanetary disc, are perfect laboratories for studying disc truncation and evolution. For the primaries, detection of dust masses of 66.6 and 32.6~M$_{\earth}$ respectively, well above debris disc level, but very low upper limits on the gas mass (a few percent of M$_{jup}$) are in stark contrast to the majority of Herbig Ae stars, which exhibit abundant gas typically above M$_{jup}$ level (Pani\'c et al., in prep). It is tempting to assign this to mutual photoevaporation - a scenario in which the discs are misaligned and allow the stars to externally photoevaporate one another. This would be very efficient in removing gas and submicron particles, leaving the larger, mm-sized dust behind. Future observations of these systems will help ascertain whether their discs are indeed misaligned.

In both systems we find the disc around the primary to be more massive and more radially extended. This may be related to the finding that disc mass scales with stellar mass, as found for single stars \citep{2016ApJ...831..125P}. We compute the expected truncation radius based on the approach of \cite{Pichardo2005CircumstellarBinaries} and find that the most likely sizes of the truncated circum-primary discs are in agreement with the fact that we are marginally resolving these discs in our observations. The wider separation pair, KK~Oph seems to be a scaled-up, higher dust mass version of the closer-separation HD~144668 pair, in spite of being older. This may suggest that dust evolution is affected by truncation much stronger than by regular disc evolution processes. 

In the case of the HD~144668 system, combining our observations with archival ALMA data at a complementary wavelength allowed us to determine that there is a large difference in opacity spectral index between the primary and secondary of $\Delta\beta=$0.69 when flux calibration is accounted for. 
We are unable to conclusively determine why this is the case. One possibility is that the secondary disc is expected to be around a factor two more compact (which would be a consequence of tidal truncation) which would correspond to more rapid dynamical processing of the dust.

\section*{Acknowledgements}
The authors thank E. Akiyama, C. Walsh, R.~Rafikov and A.~Hales for useful discussions.
The research of O.P. is funded by the Royal Society through a Royal Society Dorothy Hodgkin Fellowship. T.J.H. is also funded by a Royal Society Dorothy Hodgkin Fellowship. J.M. was funded through the Leeds University Research Scholarship. M.V. was funded through the STARRY project which received funding from the European Union's Horizon 2020 research and innovation programme under MSCA ITN-EID grant agreement No 676036. The research presented in this paper has made use of the SIMBAD database, operated at CDS, Strasbourg, France and of observations collected at the European Southern Observatory under ESO programme(s) 076.C-0708(A), 
and 095.C-0658(A). This paper makes use of the following ALMA data: 2013.0.01600.S and 2013.1.00220.S. ALMA is a partnership of ESO (representing its member states), NSF (USA) and NINS (Japan), together with NRC (Canada), MOST and ASIAA (Taiwan), and KASI (Republic of Korea), in cooperation with the Republic of Chile. The Joint ALMA Observatory is operated by ESO, AUI/NRAO and NAOJ. We thank the anonymous referee for the constructive and useful feedback.

\section{Data Availability}

All data underlying this article are available at the ALMA, ESO and GAIA Archives. 
ALMA data are found under projects 
2013.0.01600.S and 2013.1.00220.S at https://almascience.eso.org/asax/. ESO data are found under projects
076.C-0708(A)
and 095.C-0658(A) at http://archive.eso.org/cms.html.
GAIA archive can be browsed at https://gea.esac.esa.int/archive/.



\bibliography{bibliography.bib} 

\begin{thebibliography}{}
\makeatletter
\relax
\def\mn@urlcharsother{\let\do\@makeother \do\$\do\&\do\#\do\^\do\_\do\%\do\~}
\def\mn@doi{\begingroup\mn@urlcharsother \@ifnextchar [ {\mn@doi@}
  {\mn@doi@[]}}
\def\mn@doi@[#1]#2{\def\@tempa{#1}\ifx\@tempa\@empty \href
  {http://dx.doi.org/#2} {doi:#2}\else \href {http://dx.doi.org/#2} {#1}\fi
  \endgroup}
\def\mn@eprint#1#2{\mn@eprint@#1:#2::\@nil}
\def\mn@eprint@arXiv#1{\href {http://arxiv.org/abs/#1} {{\tt arXiv:#1}}}
\def\mn@eprint@dblp#1{\href {http://dblp.uni-trier.de/rec/bibtex/#1.xml}
  {dblp:#1}}
\def\mn@eprint@#1:#2:#3:#4\@nil{\def\@tempa {#1}\def\@tempb {#2}\def\@tempc
  {#3}\ifx \@tempc \@empty \let \@tempc \@tempb \let \@tempb \@tempa \fi \ifx
  \@tempb \@empty \def\@tempb {arXiv}\fi \@ifundefined
  {mn@eprint@\@tempb}{\@tempb:\@tempc}{\expandafter \expandafter \csname
  mn@eprint@\@tempb\endcsname \expandafter{\@tempc}}}

\bibitem[\protect\citeauthoryear{Alecian et~al.,}{Alecian
  et~al.}{2012}]{Alecian2012AMeasurements}
Alecian E.,  et~al., 2012, \mn@doi [Monthly Notices of the Royal Astronomical
  Society, Volume 429, Issue 2, p.1001-1026] {10.1093/mnras/sts383}, 429, 1001

\bibitem[\protect\citeauthoryear{{Andrews}}{{Andrews}}{2015}]{2015PASP..127..961A}
{Andrews} S.~M.,  2015, \mn@doi [\pasp] {10.1086/683178}, \href
  {http://adsabs.harvard.edu/abs/2015PASP..127..961A} {127, 961}

\bibitem[\protect\citeauthoryear{{Andrews} \& {Williams}}{{Andrews} \&
  {Williams}}{2005}]{2005ApJ...631.1134A}
{Andrews} S.~M.,  {Williams} J.~P.,  2005, \mn@doi [\apj] {10.1086/432712},
  \href {https://ui.adsabs.harvard.edu/abs/2005ApJ...631.1134A} {631, 1134}

\bibitem[\protect\citeauthoryear{Ansdell et~al.,}{Ansdell
  et~al.}{2016}]{Ansdell2016ALMAMasses}
Ansdell M.,  et~al., 2016, \mn@doi [The Astrophysical Journal, Volume 828,
  Issue 1, article id. 46, 15 pp. (2016).] {10.3847/0004-637X/828/1/46}, 828

\bibitem[\protect\citeauthoryear{{Baines}, {Oudmaijer}, {Porter}  \&
  {Pozzo}}{{Baines} et~al.}{2006}]{2006MNRAS.367..737B}
{Baines} D.,  {Oudmaijer} R.~D.,  {Porter} J.~M.,   {Pozzo} M.,  2006, \mn@doi
  [\mnras] {10.1111/j.1365-2966.2006.10006.x}, \href
  {https://ui.adsabs.harvard.edu/abs/2006MNRAS.367..737B} {367, 737}

\bibitem[\protect\citeauthoryear{{Birnstiel}, {Klahr}  \&
  {Ercolano}}{{Birnstiel} et~al.}{2012}]{2012A&A...539A.148B}
{Birnstiel} T.,  {Klahr} H.,   {Ercolano} B.,  2012, \mn@doi [\aap]
  {10.1051/0004-6361/201118136}, \href
  {http://adsabs.harvard.edu/abs/2012A%26A...539A.148B} {539, A148}

\bibitem[\protect\citeauthoryear{{Blondel} \& {Djie}}{{Blondel} \&
  {Djie}}{2006}]{2006A&A...456.1045B}
{Blondel} P.~F.~C.,  {Djie} H.~R.~E.~T.~A.,  2006, \mn@doi [\aap]
  {10.1051/0004-6361:20040269}, \href
  {http://adsabs.harvard.edu/abs/2006A%26A...456.1045B} {456, 1045}

\bibitem[\protect\citeauthoryear{Carmona, Van Den~Ancker  \& Henning}{Carmona
  et~al.}{2007}]{Carmona2007OpticalStars}
Carmona A.,  Van Den~Ancker M.~E.,   Henning T.,  2007, \mn@doi [A{\&}A]
  {10.1051/0004-6361:20065509}, 464, 687

\bibitem[\protect\citeauthoryear{{Carrasco-Gonz{\'a}lez}
  et~al.,}{{Carrasco-Gonz{\'a}lez} et~al.}{2016}]{2016ApJ...821L..16C}
{Carrasco-Gonz{\'a}lez} C.,  et~al., 2016, \mn@doi [\apjl]
  {10.3847/2041-8205/821/1/L16}, \href
  {https://ui.adsabs.harvard.edu/abs/2016ApJ...821L..16C} {821, L16}

\bibitem[\protect\citeauthoryear{{Cox} et~al.,}{{Cox}
  et~al.}{2017}]{2017ApJ...851...83C}
{Cox} E.~G.,  et~al., 2017, \mn@doi [\apj] {10.3847/1538-4357/aa97e2}, \href
  {http://adsabs.harvard.edu/abs/2017ApJ...851...83C} {851, 83}

\bibitem[\protect\citeauthoryear{{Cuello} et~al.,}{{Cuello}
  et~al.}{2019a}]{2019MNRAS.tmp.2545C}
{Cuello} N.,  et~al., 2019a, \mn@doi [\mnras] {10.1093/mnras/stz2938}, \href
  {https://ui.adsabs.harvard.edu/abs/2019MNRAS.tmp.2545C} {p.~2545}

\bibitem[\protect\citeauthoryear{{Cuello} et~al.,}{{Cuello}
  et~al.}{2019b}]{2019MNRAS.483.4114C}
{Cuello} N.,  et~al., 2019b, \mn@doi [\mnras] {10.1093/mnras/sty3325}, \href
  {https://ui.adsabs.harvard.edu/abs/2019MNRAS.483.4114C} {483, 4114}

\bibitem[\protect\citeauthoryear{{Dai}, {Facchini}, {Clarke}  \&
  {Haworth}}{{Dai} et~al.}{2015}]{2015MNRAS.449.1996D}
{Dai} F.,  {Facchini} S.,  {Clarke} C.~J.,   {Haworth} T.~J.,  2015, \mn@doi
  [\mnras] {10.1093/mnras/stv403}, \href
  {http://adsabs.harvard.edu/abs/2015MNRAS.449.1996D} {449, 1996}

\bibitem[\protect\citeauthoryear{{Dartois}, {Dutrey}  \&
  {Guilloteau}}{{Dartois} et~al.}{2003}]{2003A&A...399..773D}
{Dartois} E.,  {Dutrey} A.,   {Guilloteau} S.,  2003, \mn@doi [\aap]
  {10.1051/0004-6361:20021638}, \href
  {https://ui.adsabs.harvard.edu/abs/2003A&A...399..773D} {399, 773}

\bibitem[\protect\citeauthoryear{{De Rosa} et~al.,}{{De Rosa}
  et~al.}{2014}]{2014MNRAS.437.1216D}
{De Rosa} R.~J.,  et~al., 2014, \mn@doi [\mnras] {10.1093/mnras/stt1932}, \href
  {https://ui.adsabs.harvard.edu/abs/2014MNRAS.437.1216D} {437, 1216}

\bibitem[\protect\citeauthoryear{Desidera \& Barbieri}{Desidera \&
  Barbieri}{2006}]{Desidera2006PropertiesSeparation}
Desidera S.,  Barbieri M.,  2006, \mn@doi [Astronomy and Astrophysics, Volume
  462, Issue 1, January IV 2007, pp.345-353] {10.1051/0004-6361:20066319}, 462,
  345

\bibitem[\protect\citeauthoryear{{Draine}}{{Draine}}{2006}]{2006ApJ...636.1114D}
{Draine} B.~T.,  2006, \mn@doi [\apj] {10.1086/498130}, \href
  {https://ui.adsabs.harvard.edu/abs/2006ApJ...636.1114D} {636, 1114}

\bibitem[\protect\citeauthoryear{{Duch{\^e}ne}}{{Duch{\^e}ne}}{2015}]{2015Ap&SS.355..291D}
{Duch{\^e}ne} G.,  2015, \mn@doi [\apss] {10.1007/s10509-014-2173-7}, \href
  {https://ui.adsabs.harvard.edu/abs/2015Ap&SS.355..291D} {355, 291}

\bibitem[\protect\citeauthoryear{{Duch{\^e}ne} \& {Kraus}}{{Duch{\^e}ne} \&
  {Kraus}}{2013}]{2013ARA&A..51..269D}
{Duch{\^e}ne} G.,  {Kraus} A.,  2013, \mn@doi [\araa]
  {10.1146/annurev-astro-081710-102602}, \href
  {https://ui.adsabs.harvard.edu/abs/2013ARA&A..51..269D} {51, 269}

\bibitem[\protect\citeauthoryear{{Duquennoy} \& {Mayor}}{{Duquennoy} \&
  {Mayor}}{1991}]{1991A&A...248..485D}
{Duquennoy} A.,  {Mayor} M.,  1991, \aap, \href
  {https://ui.adsabs.harvard.edu/abs/1991A&A...248..485D} {500, 337}

\bibitem[\protect\citeauthoryear{{Dutrey} et~al.,}{{Dutrey}
  et~al.}{2014}]{2014Natur.514..600D}
{Dutrey} A.,  et~al., 2014, \mn@doi [\nat] {10.1038/nature13822}, \href
  {https://ui.adsabs.harvard.edu/abs/2014Natur.514..600D} {514, 600}

\bibitem[\protect\citeauthoryear{{Ercolano} \& {Pascucci}}{{Ercolano} \&
  {Pascucci}}{2017}]{2017RSOS....470114E}
{Ercolano} B.,  {Pascucci} I.,  2017, \mn@doi [Royal Society Open Science]
  {10.1098/rsos.170114}, \href
  {http://adsabs.harvard.edu/abs/2017RSOS....470114E} {4, 170114}

\bibitem[\protect\citeauthoryear{Fairlamb, Oudmaijer, Mendigut{\'{i}}a, Ilee
  \& Ancker}{Fairlamb et~al.}{2015}]{Fairlamb2015ARates}
Fairlamb J.~R.,  Oudmaijer R.~D.,  Mendigut{\'{i}}a I.,  Ilee J.~D.,   Ancker
  M. E. v.~d.,  2015, \mn@doi [Monthly Notices of the Royal Astronomical
  Society, Volume 453, Issue 1, p.976-1001] {10.1093/mnras/stv1576}, 453, 976

\bibitem[\protect\citeauthoryear{{Fontanive}, {Rice}, {Bonavita}, {Lopez},
  {Mu{\v{z}}i{\'c}}, {}  \& {Biller}}{{Fontanive}
  et~al.}{2019}]{2019MNRAS.485.4967F}
{Fontanive} C.,  {Rice} K.,  {Bonavita} M.,  {Lopez} E.,  {Mu{\v{z}}i{\'c}} {}
  K.,   {Biller} B.,  2019, \mn@doi [\mnras] {10.1093/mnras/stz671}, \href
  {https://ui.adsabs.harvard.edu/abs/2019MNRAS.485.4967F} {485, 4967}

\bibitem[\protect\citeauthoryear{{Forgan} \& {Rice}}{{Forgan} \&
  {Rice}}{2011}]{2011MNRAS.417.1928F}
{Forgan} D.,  {Rice} K.,  2011, \mn@doi [\mnras]
  {10.1111/j.1365-2966.2011.19380.x}, \href
  {https://ui.adsabs.harvard.edu/abs/2011MNRAS.417.1928F} {417, 1928}

\bibitem[\protect\citeauthoryear{{Frerking}, {Keene}, {Blake}  \&
  {Phillips}}{{Frerking} et~al.}{1989}]{1989ApJ...344..311F}
{Frerking} M.~A.,  {Keene} J.,  {Blake} G.~A.,   {Phillips} T.~G.,  1989,
  \mn@doi [\apj] {10.1086/167799}, \href
  {https://ui.adsabs.harvard.edu/abs/1989ApJ...344..311F} {344, 311}

\bibitem[\protect\citeauthoryear{{Garcia Lopez}, {Natta}, {Testi}  \&
  {Habart}}{{Garcia Lopez} et~al.}{2006}]{2006A&A...459..837G}
{Garcia Lopez} R.,  {Natta} A.,  {Testi} L.,   {Habart} E.,  2006, \mn@doi
  [\aap] {10.1051/0004-6361:20065575}, \href
  {http://adsabs.harvard.edu/abs/2006A%26A...459..837G} {459, 837}

\bibitem[\protect\citeauthoryear{{Goldreich} \& {Tremaine}}{{Goldreich} \&
  {Tremaine}}{1980}]{1980ApJ...241..425G}
{Goldreich} P.,  {Tremaine} S.,  1980, \mn@doi [\apj] {10.1086/158356}, \href
  {http://adsabs.harvard.edu/abs/1980ApJ...241..425G} {241, 425}

\bibitem[\protect\citeauthoryear{{G{\"u}ver} \& {{\"O}zel}}{{G{\"u}ver} \&
  {{\"O}zel}}{2009}]{2009MNRAS.400.2050G}
{G{\"u}ver} T.,  {{\"O}zel} F.,  2009, \mn@doi [\mnras]
  {10.1111/j.1365-2966.2009.15598.x}, \href
  {https://ui.adsabs.harvard.edu/abs/2009MNRAS.400.2050G} {400, 2050}

\bibitem[\protect\citeauthoryear{Hales et~al.,}{Hales
  et~al.}{2014}]{Hales2014AFORMATION}
Hales A.~S.,  et~al., 2014, \mn@doi [The Astronomical Journal]
  {10.1088/0004-6256/148/3/47}, 148

\bibitem[\protect\citeauthoryear{{Hall}, {Forgan}  \& {Rice}}{{Hall}
  et~al.}{2017}]{2017MNRAS.470.2517H}
{Hall} C.,  {Forgan} D.,   {Rice} K.,  2017, \mn@doi [\mnras]
  {10.1093/mnras/stx1244}, \href
  {https://ui.adsabs.harvard.edu/abs/2017MNRAS.470.2517H} {470, 2517}

\bibitem[\protect\citeauthoryear{{Hara}, {Tachihara}, {Mizuno}, {Onishi},
  {Kawamura}, {Obayashi}  \& {Fukui}}{{Hara}
  et~al.}{1999}]{1999PASJ...51..895H}
{Hara} A.,  {Tachihara} K.,  {Mizuno} A.,  {Onishi} T.,  {Kawamura} A.,
  {Obayashi} A.,   {Fukui} Y.,  1999, \mn@doi [\pasj] {10.1093/pasj/51.6.895},
  \href {https://ui.adsabs.harvard.edu/abs/1999PASJ...51..895H} {51, 895}

\bibitem[\protect\citeauthoryear{{Harries}, {Haworth}, {Acreman}, {Ali}  \&
  {Douglas}}{{Harries} et~al.}{2019}]{2019A&C....27...63H}
{Harries} T.~J.,  {Haworth} T.~J.,  {Acreman} D.,  {Ali} A.,   {Douglas} T.,
  2019, \mn@doi [Astronomy and Computing] {10.1016/j.ascom.2019.03.002}, \href
  {https://ui.adsabs.harvard.edu/abs/2019A&C....27...63H} {27, 63}

\bibitem[\protect\citeauthoryear{{Harris}, {Andrews}, {Wilner}  \&
  {Kraus}}{{Harris} et~al.}{2012a}]{2012ApJ...751..115H}
{Harris} R.~J.,  {Andrews} S.~M.,  {Wilner} D.~J.,   {Kraus} A.~L.,  2012a,
  \mn@doi [\apj] {10.1088/0004-637X/751/2/115}, \href
  {http://adsabs.harvard.edu/abs/2012ApJ...751..115H} {751, 115}

\bibitem[\protect\citeauthoryear{{Harris}, {Andrews}, {Wilner}  \&
  {Kraus}}{{Harris} et~al.}{2012b}]{Harris2012ASystems}
{Harris} R.~J.,  {Andrews} S.~M.,  {Wilner} D.~J.,   {Kraus} A.~L.,  2012b,
  \mn@doi [\apj] {10.1088/0004-637X/751/2/115}, \href
  {https://ui.adsabs.harvard.edu/abs/2012ApJ...751..115H} {751, 115}

\bibitem[\protect\citeauthoryear{{Hartmann}, {Calvet}, {Gullbring}  \&
  {D'Alessio}}{{Hartmann} et~al.}{1998}]{1998ApJ...495..385H}
{Hartmann} L.,  {Calvet} N.,  {Gullbring} E.,   {D'Alessio} P.,  1998, \mn@doi
  [\apj] {10.1086/305277}, \href
  {https://ui.adsabs.harvard.edu/abs/1998ApJ...495..385H} {495, 385}

\bibitem[\protect\citeauthoryear{{Haworth} et~al.,}{{Haworth}
  et~al.}{2016}]{2016PASA...33...53H}
{Haworth} T.~J.,  et~al., 2016, \mn@doi [\pasa] {10.1017/pasa.2016.45}, \href
  {http://adsabs.harvard.edu/abs/2016PASA...33...53H} {33, e053}

\bibitem[\protect\citeauthoryear{{Herbig}}{{Herbig}}{1960}]{1960ApJS....4..337H}
{Herbig} G.~H.,  1960, \mn@doi [\apjs] {10.1086/190050}, \href
  {http://adsabs.harvard.edu/abs/1960ApJS....4..337H} {4, 337}

\bibitem[\protect\citeauthoryear{{Herbig}}{{Herbig}}{2005}]{2005AJ....130..815H}
{Herbig} G.~H.,  2005, \mn@doi [\aj] {10.1086/431759}, \href
  {http://adsabs.harvard.edu/abs/2005AJ....130..815H} {130, 815}

\bibitem[\protect\citeauthoryear{Hillenbrand, Strom, Vrba  \&
  Keene}{Hillenbrand et~al.}{1992}]{Hillenbrand1992HerbigDisks}
Hillenbrand L.~A.,  Strom S.~E.,  Vrba F.~J.,   Keene J.,  1992, \mn@doi [The
  Astrophysical Journal] {10.1086/171819}, 397, 613

\bibitem[\protect\citeauthoryear{{Jaeger}}{{Jaeger}}{2008}]{2008ASPC..394..623J}
{Jaeger} S.,  2008, in {Argyle} R.~W.,  {Bunclark} P.~S.,   {Lewis} J.~R.,
  eds,  Astronomical Society of the Pacific Conference Series Vol. 394,
  Astronomical Data Analysis Software and Systems XVII. p.~623

\bibitem[\protect\citeauthoryear{{Jankovic}, {Owen}  \& {Mohanty}}{{Jankovic}
  et~al.}{2019}]{2019MNRAS.484.2296J}
{Jankovic} M.~R.,  {Owen} J.~E.,   {Mohanty} S.,  2019, \mn@doi [\mnras]
  {10.1093/mnras/stz004}, \href
  {https://ui.adsabs.harvard.edu/abs/2019MNRAS.484.2296J} {484, 2296}

\bibitem[\protect\citeauthoryear{{Jensen}, {Mathieu}  \& {Fuller}}{{Jensen}
  et~al.}{1994}]{1994ApJ...429L..29J}
{Jensen} E. L.~N.,  {Mathieu} R.~D.,   {Fuller} G.~A.,  1994, \mn@doi [\apjl]
  {10.1086/187405}, \href
  {https://ui.adsabs.harvard.edu/abs/1994ApJ...429L..29J} {429, L29}

\bibitem[\protect\citeauthoryear{Kennedy et~al.,}{Kennedy
  et~al.}{2019}]{Kennedy2019}
Kennedy G.~M.,  et~al., 2019, \mn@doi [Nature Astronomy]
  {10.1038/s41550-018-0667-x}, 3, 230

\bibitem[\protect\citeauthoryear{{Kouwenhoven}, {Goodwin}, {Parker}, {Davies},
  {Malmberg}  \& {Kroupa}}{{Kouwenhoven} et~al.}{2010}]{2010MNRAS.404.1835K}
{Kouwenhoven} M.~B.~N.,  {Goodwin} S.~P.,  {Parker} R.~J.,  {Davies} M.~B.,
  {Malmberg} D.,   {Kroupa} P.,  2010, \mn@doi [\mnras]
  {10.1111/j.1365-2966.2010.16399.x}, \href
  {https://ui.adsabs.harvard.edu/abs/2010MNRAS.404.1835K} {404, 1835}

\bibitem[\protect\citeauthoryear{{Kratter}, {Matzner}, {Krumholz}  \&
  {Klein}}{{Kratter} et~al.}{2010}]{2010ApJ...708.1585K}
{Kratter} K.~M.,  {Matzner} C.~D.,  {Krumholz} M.~R.,   {Klein} R.~I.,  2010,
  \mn@doi [\apj] {10.1088/0004-637X/708/2/1585}, \href
  {https://ui.adsabs.harvard.edu/abs/2010ApJ...708.1585K} {708, 1585}

\bibitem[\protect\citeauthoryear{{Kraus}, {Ireland}, {Huber}, {Mann}  \&
  {Dupuy}}{{Kraus} et~al.}{2016}]{2016AJ....152....8K}
{Kraus} A.~L.,  {Ireland} M.~J.,  {Huber} D.,  {Mann} A.~W.,   {Dupuy} T.~J.,
  2016, \mn@doi [\aj] {10.3847/0004-6256/152/1/8}, \href
  {https://ui.adsabs.harvard.edu/abs/2016AJ....152....8K} {152, 8}

\bibitem[\protect\citeauthoryear{{Kreplin} et~al.,}{{Kreplin}
  et~al.}{2013}]{2013A&A...551A..21K}
{Kreplin} A.,  et~al., 2013, \mn@doi [\aap] {10.1051/0004-6361/201220806},
  \href {http://adsabs.harvard.edu/abs/2013A%26A...551A..21K} {551, A21}

\bibitem[\protect\citeauthoryear{{Kurtovic} et~al.,}{{Kurtovic}
  et~al.}{2018}]{2018ApJ...869L..44K}
{Kurtovic} N.~T.,  et~al., 2018, \mn@doi [\apjl] {10.3847/2041-8213/aaf746},
  \href {https://ui.adsabs.harvard.edu/abs/2018ApJ...869L..44K} {869, L44}

\bibitem[\protect\citeauthoryear{{Lallement}, {Babusiaux}, {Vergely}, {Katz},
  {Arenou}, {Valette}, {Hottier}  \& {Capitanio}}{{Lallement}
  et~al.}{2019}]{2019A&A...625A.135L}
{Lallement} R.,  {Babusiaux} C.,  {Vergely} J.~L.,  {Katz} D.,  {Arenou} F.,
  {Valette} B.,  {Hottier} C.,   {Capitanio} L.,  2019, \mn@doi [\aap]
  {10.1051/0004-6361/201834695}, \href
  {https://ui.adsabs.harvard.edu/abs/2019A&A...625A.135L} {625, A135}

\bibitem[\protect\citeauthoryear{{Leinert}, {Richichi}  \& {Haas}}{{Leinert}
  et~al.}{1997}]{1997A&A...318..472L}
{Leinert} C.,  {Richichi} A.,   {Haas} M.,  1997, \aap, \href
  {http://adsabs.harvard.edu/abs/1997A%26A...318..472L} {318, 472}

\bibitem[\protect\citeauthoryear{{Leinert} et~al.,}{{Leinert}
  et~al.}{2004}]{2004A&A...423..537L}
{Leinert} C.,  et~al., 2004, \mn@doi [\aap] {10.1051/0004-6361:20047178}, \href
  {http://adsabs.harvard.edu/abs/2004A%26A...423..537L} {423, 537}

\bibitem[\protect\citeauthoryear{{Lombardi}, {Alves}  \& {Lada}}{{Lombardi}
  et~al.}{2006}]{2006A&A...454..781L}
{Lombardi} M.,  {Alves} J.,   {Lada} C.~J.,  2006, \mn@doi [\aap]
  {10.1051/0004-6361:20042474}, \href
  {https://ui.adsabs.harvard.edu/abs/2006A&A...454..781L} {454, 781}

\bibitem[\protect\citeauthoryear{{Lombardi}, {Lada}  \& {Alves}}{{Lombardi}
  et~al.}{2008}]{2008A&A...489..143L}
{Lombardi} M.,  {Lada} C.~J.,   {Alves} J.,  2008, \mn@doi [\aap]
  {10.1051/0004-6361:200810070}, \href
  {https://ui.adsabs.harvard.edu/abs/2008A&A...489..143L} {489, 143}

\bibitem[\protect\citeauthoryear{{Long} et~al.,}{{Long}
  et~al.}{2018}]{2018ApJ...863...61L}
{Long} F.,  et~al., 2018, \mn@doi [\apj] {10.3847/1538-4357/aacce9}, \href
  {https://ui.adsabs.harvard.edu/abs/2018ApJ...863...61L} {863, 61}

\bibitem[\protect\citeauthoryear{{Manara} et~al.,}{{Manara}
  et~al.}{2016}]{2016A&A...591L...3M}
{Manara} C.~F.,  et~al., 2016, \mn@doi [\aap] {10.1051/0004-6361/201628549},
  \href {https://ui.adsabs.harvard.edu/abs/2016A&A...591L...3M} {591, L3}

\bibitem[\protect\citeauthoryear{{Manara}, {Morbidelli}  \& {Guillot}}{{Manara}
  et~al.}{2018}]{2018A&A...618L...3M}
{Manara} C.~F.,  {Morbidelli} A.,   {Guillot} T.,  2018, \mn@doi [\aap]
  {10.1051/0004-6361/201834076}, \href
  {https://ui.adsabs.harvard.edu/abs/2018A&A...618L...3M} {618, L3}

\bibitem[\protect\citeauthoryear{{Manara} et~al.,}{{Manara}
  et~al.}{2019}]{2019A&A...628A..95M}
{Manara} C.~F.,  et~al., 2019, \mn@doi [\aap] {10.1051/0004-6361/201935964},
  \href {https://ui.adsabs.harvard.edu/abs/2019A&A...628A..95M} {628, A95}

\bibitem[\protect\citeauthoryear{{Mann} \& {Williams}}{{Mann} \&
  {Williams}}{2009}]{2009ApJ...699L..55M}
{Mann} R.~K.,  {Williams} J.~P.,  2009, \mn@doi [\apjl]
  {10.1088/0004-637X/699/1/L55}, \href
  {https://ui.adsabs.harvard.edu/abs/2009ApJ...699L..55M} {699, L55}

\bibitem[\protect\citeauthoryear{{Matr{\`a}}, {Pani{\'c}}, {Wyatt}  \&
  {Dent}}{{Matr{\`a}} et~al.}{2015}]{2015MNRAS.447.3936M}
{Matr{\`a}} L.,  {Pani{\'c}} O.,  {Wyatt} M.~C.,   {Dent} W.~R.~F.,  2015,
  \mn@doi [\mnras] {10.1093/mnras/stu2619}, \href
  {https://ui.adsabs.harvard.edu/abs/2015MNRAS.447.3936M} {447, 3936}

\bibitem[\protect\citeauthoryear{{Mayama} et~al.,}{{Mayama}
  et~al.}{2010}]{2010Sci...327..306M}
{Mayama} S.,  et~al., 2010, \mn@doi [Science] {10.1126/science.1179679}, \href
  {https://ui.adsabs.harvard.edu/abs/2010Sci...327..306M} {327, 306}

\bibitem[\protect\citeauthoryear{{McCarthy} \& {Zuckerman}}{{McCarthy} \&
  {Zuckerman}}{2004}]{2004AJ....127.2871M}
{McCarthy} C.,  {Zuckerman} B.,  2004, \mn@doi [\aj] {10.1086/383559}, \href
  {https://ui.adsabs.harvard.edu/abs/2004AJ....127.2871M} {127, 2871}

\bibitem[\protect\citeauthoryear{Meeus et~al.,}{Meeus
  et~al.}{2012a}]{Meeus2012AstrophysicsDiscs}
Meeus G.,  et~al., 2012a, \mn@doi [A{\&}A] {10.1051/0004-6361/201219225}, 544

\bibitem[\protect\citeauthoryear{{Meeus} et~al.,}{{Meeus}
  et~al.}{2012b}]{Meeus2012GASPSDiscs}
{Meeus} G.,  et~al., 2012b, \mn@doi [\aap] {10.1051/0004-6361/201219225}, \href
  {https://ui.adsabs.harvard.edu/abs/2012A&A...544A..78M} {544, A78}

\bibitem[\protect\citeauthoryear{{Miley}, {Pani{\'c}}, {Wyatt}  \&
  {Kennedy}}{{Miley} et~al.}{2018}]{2018A&A...615L..10M}
{Miley} J.~M.,  {Pani{\'c}} O.,  {Wyatt} M.,   {Kennedy} G.~M.,  2018, \mn@doi
  [\aap] {10.1051/0004-6361/201833381}, \href
  {https://ui.adsabs.harvard.edu/abs/2018A&A...615L..10M} {615, L10}

\bibitem[\protect\citeauthoryear{{Miley}, {Pani{\'c}}, {Haworth}, {Pascucci},
  {Wyatt}, {Clarke}, {Richards}  \& {Ratzka}}{{Miley}
  et~al.}{2019}]{2019MNRAS.485..739M}
{Miley} J.~M.,  {Pani{\'c}} O.,  {Haworth} T.~J.,  {Pascucci} I.,  {Wyatt} M.,
  {Clarke} C.,  {Richards} A.~M.~S.,   {Ratzka} T.,  2019, \mn@doi [\mnras]
  {10.1093/mnras/stz426}, \href
  {https://ui.adsabs.harvard.edu/abs/2019MNRAS.485..739M} {485, 739}

\bibitem[\protect\citeauthoryear{{Miotello}, {Bruderer}  \& {van
  Dishoeck}}{{Miotello} et~al.}{2014}]{2014A&A...572A..96M}
{Miotello} A.,  {Bruderer} S.,   {van Dishoeck} E.~F.,  2014, \mn@doi [\aap]
  {10.1051/0004-6361/201424712}, \href
  {http://adsabs.harvard.edu/abs/2014A%26A...572A..96M} {572, A96}

\bibitem[\protect\citeauthoryear{Miotello et~al.,}{Miotello
  et~al.}{2016}]{Miotello2016LupusDepletion}
Miotello A.,  et~al., 2016, \mn@doi [Astronomy {\&} Astrophysics, Volume 599,
  id.A113, 10 pp.] {10.1051/0004-6361/201629556}, 599

\bibitem[\protect\citeauthoryear{{M{\"u}ller}, {Thorwirth}, {Roth}  \&
  {Winnewisser}}{{M{\"u}ller} et~al.}{2001}]{2001A&A...370L..49M}
{M{\"u}ller} H.~S.~P.,  {Thorwirth} S.,  {Roth} D.~A.,   {Winnewisser} G.,
  2001, \mn@doi [\aap] {10.1051/0004-6361:20010367}, \href
  {https://ui.adsabs.harvard.edu/abs/2001A&A...370L..49M} {370, L49}

\bibitem[\protect\citeauthoryear{{Ogilvie} \& {Dubus}}{{Ogilvie} \&
  {Dubus}}{2001}]{10.1046/j.1365-8711.2001.04011.x}
{Ogilvie} G.~I.,  {Dubus} G.,  2001, \mn@doi [\mnras]
  {10.1046/j.1365-8711.2001.04011.x}, \href
  {https://ui.adsabs.harvard.edu/abs/2001MNRAS.320..485O} {320, 485}

\bibitem[\protect\citeauthoryear{{Onishi} et~al.,}{{Onishi}
  et~al.}{1999}]{1999PASJ...51..871O}
{Onishi} T.,  et~al., 1999, \mn@doi [\pasj] {10.1093/pasj/51.6.871}, \href
  {https://ui.adsabs.harvard.edu/abs/1999PASJ...51..871O} {51, 871}

\bibitem[\protect\citeauthoryear{{Owen} \& {Lai}}{{Owen} \&
  {Lai}}{2017}]{2017MNRAS.469.2834O}
{Owen} J.~E.,  {Lai} D.,  2017, \mn@doi [\mnras] {10.1093/mnras/stx1033}, \href
  {https://ui.adsabs.harvard.edu/abs/2017MNRAS.469.2834O} {469, 2834}

\bibitem[\protect\citeauthoryear{{Papaloizou} \& {Pringle}}{{Papaloizou} \&
  {Pringle}}{1977}]{1977MNRAS.181..441P}
{Papaloizou} J.,  {Pringle} J.~E.,  1977, \mn@doi [\mnras]
  {10.1093/mnras/181.3.441}, \href
  {https://ui.adsabs.harvard.edu/abs/1977MNRAS.181..441P} {181, 441}

\bibitem[\protect\citeauthoryear{{Papaloizou} \& {Pringle}}{{Papaloizou} \&
  {Pringle}}{1983}]{1983MNRAS.202.1181P}
{Papaloizou} J.~C.~B.,  {Pringle} J.~E.,  1983, \mn@doi [\mnras]
  {10.1093/mnras/202.4.1181}, \href
  {http://adsabs.harvard.edu/abs/1983MNRAS.202.1181P} {202, 1181}

\bibitem[\protect\citeauthoryear{Pascual, Montesinos, Meeus, Marshall,
  Mendigut{\'{i}}a  \& Sandell}{Pascual
  et~al.}{2016}]{Pascual2016ThePhotometry}
Pascual N.,  Montesinos B.,  Meeus G.,  Marshall J.~P.,  Mendigut{\'{i}}a I.,
  Sandell G.,  2016, \mn@doi [A{\&}A] {10.1051/0004-6361/201526605}, 586

\bibitem[\protect\citeauthoryear{{Pascucci} et~al.,}{{Pascucci}
  et~al.}{2016}]{2016ApJ...831..125P}
{Pascucci} I.,  et~al., 2016, \mn@doi [\apj] {10.3847/0004-637X/831/2/125},
  \href {https://ui.adsabs.harvard.edu/abs/2016ApJ...831..125P} {831, 125}

\bibitem[\protect\citeauthoryear{Pichardo, Sparke  \& Aguilar}{Pichardo
  et~al.}{2005}]{Pichardo2005CircumstellarBinaries}
Pichardo B.,  Sparke L.~S.,   Aguilar L.~A.,  2005, \mn@doi [Monthly Notices of
  the Royal Astronomical Society, Volume 359, Issue 2, pp. 521-530.]
  {10.1111/j.1365-2966.2005.08905.x}, 359, 521

\bibitem[\protect\citeauthoryear{{Pickett}, {Poynter}, {Cohen}, {Delitsky},
  {Pearson}  \& {M{\"u}ller}}{{Pickett} et~al.}{1998}]{1998JQSRT..60..883P}
{Pickett} H.~M.,  {Poynter} R.~L.,  {Cohen} E.~A.,  {Delitsky} M.~L.,
  {Pearson} J.~C.,   {M{\"u}ller} H.~S.~P.,  1998, \mn@doi [\jqsrt]
  {10.1016/S0022-4073(98)00091-0}, \href
  {https://ui.adsabs.harvard.edu/abs/1998JQSRT..60..883P} {60, 883}

\bibitem[\protect\citeauthoryear{{Pi{\'e}tu}, {Dutrey}  \&
  {Guilloteau}}{{Pi{\'e}tu} et~al.}{2007}]{2007A&A...467..163P}
{Pi{\'e}tu} V.,  {Dutrey} A.,   {Guilloteau} S.,  2007, \mn@doi [\aap]
  {10.1051/0004-6361:20066537}, \href
  {https://ui.adsabs.harvard.edu/abs/2007A&A...467..163P} {467, 163}

\bibitem[\protect\citeauthoryear{{Pirzkal}, {Spillar}  \& {Dyck}}{{Pirzkal}
  et~al.}{1997}]{1997ApJ...481..392P}
{Pirzkal} N.,  {Spillar} E.~J.,   {Dyck} H.~M.,  1997, \mn@doi [\apj]
  {10.1086/304055}, \href {http://adsabs.harvard.edu/abs/1997ApJ...481..392P}
  {481, 392}

\bibitem[\protect\citeauthoryear{{Rafikov}}{{Rafikov}}{2006}]{2006ApJ...648..666R}
{Rafikov} R.~R.,  2006, \mn@doi [\apj] {10.1086/505695}, \href
  {https://ui.adsabs.harvard.edu/abs/2006ApJ...648..666R} {648, 666}

\bibitem[\protect\citeauthoryear{{Raghavan}, {Henry}, {Mason}, {Subasavage},
  {Jao}, {Beaulieu}  \& {Hambly}}{{Raghavan}
  et~al.}{2006}]{2006ApJ...646..523R}
{Raghavan} D.,  {Henry} T.~J.,  {Mason} B.~D.,  {Subasavage} J.~P.,  {Jao}
  W.-C.,  {Beaulieu} T.~D.,   {Hambly} N.~C.,  2006, \mn@doi [\apj]
  {10.1086/504823}, \href
  {https://ui.adsabs.harvard.edu/abs/2006ApJ...646..523R} {646, 523}

\bibitem[\protect\citeauthoryear{{Reffert}, {Bergmann}, {Quirrenbach},
  {Trifonov}  \& {K{\"u}nstler}}{{Reffert} et~al.}{2015}]{2015A&A...574A.116R}
{Reffert} S.,  {Bergmann} C.,  {Quirrenbach} A.,  {Trifonov} T.,
  {K{\"u}nstler} A.,  2015, \mn@doi [\aap] {10.1051/0004-6361/201322360}, \href
  {https://ui.adsabs.harvard.edu/abs/2015A&A...574A.116R} {574, A116}

\bibitem[\protect\citeauthoryear{{Ricci}, {Testi}, {Williams}, {Mann}  \&
  {Birnstiel}}{{Ricci} et~al.}{2011}]{2011ApJ...739L...8R}
{Ricci} L.,  {Testi} L.,  {Williams} J.~P.,  {Mann} R.~K.,   {Birnstiel} T.,
  2011, \mn@doi [\apjl] {10.1088/2041-8205/739/1/L8}, \href
  {https://ui.adsabs.harvard.edu/abs/2011ApJ...739L...8R} {739, L8}

\bibitem[\protect\citeauthoryear{{Richards}}{{Richards}}{1997}]{1997PhDT........20R}
{Richards} A.~M.~S.,  1997, PhD thesis, -

\bibitem[\protect\citeauthoryear{{Rodriguez} \& {Zuckerman}}{{Rodriguez} \&
  {Zuckerman}}{2012}]{2012ApJ...745..147R}
{Rodriguez} D.~R.,  {Zuckerman} B.,  2012, \mn@doi [\apj]
  {10.1088/0004-637X/745/2/147}, \href
  {https://ui.adsabs.harvard.edu/abs/2012ApJ...745..147R} {745, 147}

\bibitem[\protect\citeauthoryear{{Sch{\"o}ier}, {van der Tak}, {van Dishoeck}
  \& {Black}}{{Sch{\"o}ier} et~al.}{2005}]{2005A&A...432..369S}
{Sch{\"o}ier} F.~L.,  {van der Tak} F.~F.~S.,  {van Dishoeck} E.~F.,   {Black}
  J.~H.,  2005, \mn@doi [\aap] {10.1051/0004-6361:20041729}, \href
  {https://ui.adsabs.harvard.edu/abs/2005A&A...432..369S} {432, 369}

\bibitem[\protect\citeauthoryear{{Schultz} \& {Wiemer}}{{Schultz} \&
  {Wiemer}}{1975}]{1975A&A....43..133S}
{Schultz} G.~V.,  {Wiemer} W.,  1975, \aap, \href
  {https://ui.adsabs.harvard.edu/abs/1975A&A....43..133S} {43, 133}

\bibitem[\protect\citeauthoryear{{Stamatellos} \& {Whitworth}}{{Stamatellos} \&
  {Whitworth}}{2008}]{2008A&A...480..879S}
{Stamatellos} D.,  {Whitworth} A.~P.,  2008, \mn@doi [\aap]
  {10.1051/0004-6361:20078628}, \href
  {https://ui.adsabs.harvard.edu/abs/2008A&A...480..879S} {480, 879}

\bibitem[\protect\citeauthoryear{{Testi} et~al.,}{{Testi}
  et~al.}{2014}]{2014prpl.conf..339T}
{Testi} L.,  et~al., 2014, in {Beuther} H.,  {Klessen} R.~S.,  {Dullemond}
  C.~P.,   {Henning} T.,  eds, Protostars and Planets VI. p.~339 (\mn@eprint
  {arXiv} {1402.1354}), \mn@doi{10.2458/azu_uapress_9780816531240-ch015}

\bibitem[\protect\citeauthoryear{{Thompson}, {Moran}  \& {Swenson}}{{Thompson}
  et~al.}{2017}]{2017isra.book.....T}
{Thompson} A.~R.,  {Moran} J.~M.,   {Swenson} George~W. J.,  2017,
  {Interferometry and Synthesis in Radio Astronomy, 3rd Edition},
  \mn@doi{10.1007/978-3-319-44431-4.
}

\bibitem[\protect\citeauthoryear{{Thorngren}, {Fortney}, {Murray-Clay}  \&
  {Lopez}}{{Thorngren} et~al.}{2016}]{2016ApJ...831...64T}
{Thorngren} D.~P.,  {Fortney} J.~J.,  {Murray-Clay} R.~A.,   {Lopez} E.~D.,
  2016, \mn@doi [\apj] {10.3847/0004-637X/831/1/64}, \href
  {https://ui.adsabs.harvard.edu/abs/2016ApJ...831...64T} {831, 64}

\bibitem[\protect\citeauthoryear{Torres \& {Guillermo}}{Torres \&
  {Guillermo}}{1999}]{Torres1999SubstellarApproach}
Torres G.,  {Guillermo} 1999, \mn@doi [Publications of the Astronomical Society
  of the Pacific] {10.1086/316313}, 111, 169

\bibitem[\protect\citeauthoryear{{Tremaine} \& {Davis}}{{Tremaine} \&
  {Davis}}{2014}]{10.1093/mnras/stu663}
{Tremaine} S.,  {Davis} S.~W.,  2014, \mn@doi [\mnras] {10.1093/mnras/stu663},
  \href {https://ui.adsabs.harvard.edu/abs/2014MNRAS.441.1408T} {441, 1408}

\bibitem[\protect\citeauthoryear{{Tripathi}, {Andrews}, {Birnstiel}  \&
  {Wilner}}{{Tripathi} et~al.}{2017}]{2017ApJ...845...44T}
{Tripathi} A.,  {Andrews} S.~M.,  {Birnstiel} T.,   {Wilner} D.~J.,  2017,
  \mn@doi [\apj] {10.3847/1538-4357/aa7c62}, \href
  {https://ui.adsabs.harvard.edu/abs/2017ApJ...845...44T} {845, 44}

\bibitem[\protect\citeauthoryear{{Vioque}, {Oudmaijer}, {Baines},
  {Mendigut{\'\i}a}  \& {P{\'e}rez-Mart{\'\i}nez}}{{Vioque}
  et~al.}{2018}]{2018A&A...620A.128V}
{Vioque} M.,  {Oudmaijer} R.~D.,  {Baines} D.,  {Mendigut{\'\i}a} I.,
  {P{\'e}rez-Mart{\'\i}nez} R.,  2018, \mn@doi [\aap]
  {10.1051/0004-6361/201832870}, \href
  {https://ui.adsabs.harvard.edu/abs/2018A&A...620A.128V} {620, A128}

\bibitem[\protect\citeauthoryear{Wenger et~al.,}{Wenger
  et~al.}{2000}]{Wenger2000TheDatabase}
Wenger M.,  et~al., 2000, \mn@doi [Astronomy and Astrophysics Supplement,
  v.143, p.9-22] {10.1051/aas:2000332}, 143, 9

\bibitem[\protect\citeauthoryear{{Williams} \& {Cieza}}{{Williams} \&
  {Cieza}}{2011}]{2011ARA&A..49...67W}
{Williams} J.~P.,  {Cieza} L.~A.,  2011, \mn@doi [\araa]
  {10.1146/annurev-astro-081710-102548}, \href
  {http://adsabs.harvard.edu/abs/2011ARA%26A..49...67W} {49, 67}

\bibitem[\protect\citeauthoryear{{Wilson} \& {Rood}}{{Wilson} \&
  {Rood}}{1994}]{1994ARA&A..32..191W}
{Wilson} T.~L.,  {Rood} R.,  1994, \mn@doi [\araa]
  {10.1146/annurev.aa.32.090194.001203}, \href
  {https://ui.adsabs.harvard.edu/abs/1994ARA&A..32..191W} {32, 191}

\bibitem[\protect\citeauthoryear{{Winn} \& {Fabrycky}}{{Winn} \&
  {Fabrycky}}{2015}]{2015ARA&A..53..409W}
{Winn} J.~N.,  {Fabrycky} D.~C.,  2015, \mn@doi [\araa]
  {10.1146/annurev-astro-082214-122246}, \href
  {http://adsabs.harvard.edu/abs/2015ARA%26A..53..409W} {53, 409}

\bibitem[\protect\citeauthoryear{{Woitke} et~al.,}{{Woitke}
  et~al.}{2016}]{2016A&A...586A.103W}
{Woitke} P.,  et~al., 2016, \mn@doi [\aap] {10.1051/0004-6361/201526538}, \href
  {https://ui.adsabs.harvard.edu/abs/2016A&A...586A.103W} {586, A103}

\bibitem[\protect\citeauthoryear{Zucker \& Mazeh}{Zucker \&
  Mazeh}{2002}]{Zucker2002ONPLANETS}
Zucker S.,  Mazeh T.,  2002, The Astrophysical Journal, 568, 113

\makeatother
\end{thebibliography}




\bsp	
\label{lastpage}
\end{document}